%% file: main.tex
\documentclass[11pt]{article}
\usepackage[ruled,vlined]{algorithm2e}
\usepackage{macros}
\usepackage{doi}
\usepackage[dvipsnames]{xcolor}
\hypersetup{
	colorlinks=true,
	urlcolor=MidnightBlue,
	linkcolor=MidnightBlue,
	citecolor=MidnightBlue,
	unicode
}
\usepackage{microtype}
\usepackage{yhmath}
\usepackage{braket}
\usepackage{cancel}
\usepackage[backend=biber,maxbibnames=10,maxalphanames=4, style=alphabetic]{biblatex}
\DeclareFieldFormat{titlecase}{#1}
\usepackage{authblk}

\title{Compressed Oracle Duality: A simple framework \\for knowing what quantum adversaries know}

\author{Matthew Ding\thanks{\texttt{mcding@stanford.edu}} }
\author{Mark Zhandry\thanks{\texttt{mzhandry@stanford.edu}}}
\affil{Department of Computer Science, Stanford University}
\date{October 30, 2026}
\begin{document}

\maketitle

\begin{abstract}
    Given a quantum adversary that queries a random oracle, how do we characterize the information that they ``know''? The Compressed Oracle method of Zhandry (CRYPTO 2019) is one method that records the queries that the adversary makes; this recording serves as a representation of the information that is learned. However, a significant limitation of the Compressed Oracle method is that it may continue to store information that the adversary has ``forgotten.'' This issue is especially pertinent in the space-bounded setting, where the adversary may uncompute information to free up space, while the information may still reside in the compressed oracle. 
    
    In this work, we present a simple and interpretable framework called ``Compressed Oracle Duality,'' which extends the Compressed Oracle method for space-bounded algorithms. By applying an isometry to the memory of a quantum adversary, we obtain a pair of compressed oracles with identical marginal distributions and classical correlations which correspond to the knowledge the adversary retains. In fact, we show that the outcome of measuring the new compressed oracle database corresponds exactly to applying the pretty-good measurement on the adversary's memory.
    
    As an application of our framework for collision search, we give a simplified proof of the recent $T^2S = \Omega(N\log N)$ quantum time-space tradeoff for label-symmetric algorithms due to Magniez--Zur (arXiv:2609.10808); we also prove a stronger \emph{classical} tradeoff of $TS=\Omega(N\log N)$ assuming label symmetry. The proofs under our framework do not require representation theory and instead bound the mutual information between the compressed oracle pairs. This highlights both how the label symmetry assumption bypasses known proof barriers as well as the challenges that remain in achieving unconditional quantum time-space tradeoffs.
\end{abstract}

\newpage
\section{Introduction}

Quantum query complexity is one of the main testbeds for understanding the power of quantum computation. The special case of a uniformly random oracle as an input is also especially important for cryptography: most practical cryptosystems utilize a cryptographic hash function, which in turn is typically modeled as a random oracle~\cite{BR93}. The security of such systems then relies on proving strong query lower bounds for random oracles. In the quantum setting, one must therefore prove quantum query lower bounds~\cite{Boneh_2011}.

Despite much progress over the last thirty years in proving quantum query lower bounds, many open problems remain. One of the main challenges is to understand what kind of information, and how much, an adversary can ``know'' after querying the oracle a limited number of times. 

\paragraph{This work.} We introduce \emph{Compressed Oracle Duality}, a framework for analyzing the information a quantum adversary retains about a random oracle. We start with the Compressed Oracle method~\cite{zhandry2019record}, which purifies the random oracle and re-encodes it as a small database $D$ of input/output pairs. We then take the additional step of canonically re-encoding the \emph{adversary’s} memory as a second compressed oracle $\tilde{D}$, which has the same database distribution as the original. Correlations between measurement outcomes on $\tilde{D}$ and $D$ give quantitative bounds on information recoverable from memory without further queries. 

As an application, we give a simpler proof of the very recent $T^2S=\Omega(N\log N)$ collision-finding lower bound for ``label-symmetric'' quantum algorithms due to Magniez and Zur~\cite{magniez2026label}, replacing representation-theoretic arguments with information-theoretic bounds; we also prove a classical lower bound of $TS=\Omega(N\log N)$ under label symmetry using a similar technique.\footnote{We stress that Compressed Oracle Duality does not rely on label symmetry; only these particular applications utilize label symmetry.} Pollard's rho algorithm has a smaller time-space product of $TS=\Os(N^{1/2}\log N)$, showing that the label symmetry restriction excludes successful classical low-space strategies. This demonstrates a significant obstacle to extending lower bound strategies based on label symmetry to unrestricted quantum algorithms.

\subsection{Motivation: Quantum time-space lower bounds for collision-finding} A key open quantum query complexity question, which will also be the main task studied in this work, is to understand the complexity of computing collisions in a random function $\Hm:[M]\rightarrow[N]$, $M\geq N$, given a limited amount of storage capacity. Ignoring space, the quantum query complexity is known to be $\Theta(N^{1/3})$~\cite{Brassard_1998,aaronsonshi2004,Zhandry_2015}\footnote{\cite{Brassard_1998} showed the upper bound and~\cite{aaronsonshi2004} showed the lower bound for the case of worst-case/random 2-to-1 functions. For uniformly random functions, the algorithm of~\cite{Brassard_1998} readily applies, and the lower bound was proved by~\cite{Zhandry_2015}.}. However, the algorithm of Brassard--H{\o}yer--Tapp~\cite{Brassard_1998}, henceforth BHT, also requires space $S=\Omega(N^{1/3}\log N)$. While BHT is readily adapted to use smaller space $S$, this comes at the cost of increasing the number of queries $T$, resulting in a general tradeoff of $T^2S=\Os(N\log N)$. Even after almost thirty years, this remains the best known quantum algorithm for collision-finding; on the other hand, no non-trivial unrestricted space lower bounds are known.

It is natural to conjecture that $T^2S=\Theta(N\log N)$ is optimal. However, the classical setting may actually suggest otherwise: the classical query complexity of finding collisions is $T=\Theta(N^{1/2})$. The simple ``birthday attack'' uses space $S=\Os(N^{1/2}\log N)$, and a natural space-limited version achieves a general tradeoff of $TS=\Os(N\log N)$. This appears to parallel what is known for quantum algorithms, just with the dependence on $T$ differing in the classical versus quantum settings. Indeed, the space-limited BHT algorithm is a natural analog of the space-limited birthday attack. However, a clever classical algorithm known as Pollard's rho~\cite{pollardrho1975,vanoorschot1999} enables setting $T=\Os(N^{1/2})$ and $S=\Os(\log N)$, simultaneously achieving optimal query and space complexity. As such, one may expect that similar ideas to Pollard's rho may lead to a low-space quantum algorithm achieving $T=\Os(N^{1/3})$ and $S=\Os(\log N)$. However, no such quantum algorithm has been discovered. Our motivating question is therefore:

\begin{center}
    \textbf{\emph{What is the optimal time-space tradeoff for quantum collision-finding?}}
\end{center}

This question has circulated for more than a decade, and was highlighted in Aaronson's list of twelve open problems in quantum query complexity~\cite{Aaronson2021open}. In fact, quantum time-space tradeoffs for collision-finding remain arguably the only question on the list without a claimed full solution.\footnote{Solutions for several of these problems were generated very recently with significant LLM usage, and many of these have not yet been peer reviewed.}

Moreover, collision time-space tradeoffs also matter for cryptographic parameter choices: cryptographic hash functions will often set $N=2^{256}$, corresponding to a 128-bit security level assuming the optimality of classical birthday attacks.\footnote{The security level is the log (base 2) of run-time of the fastest attack.} Considering the random-oracle query model and ignoring space, BHT would suggest that we need $N\geq 2^{384}$ against quantum attacks, resulting in less efficient overall systems. However, it is widely believed that the enormous space of BHT means $N=2^{384}$ is overly conservative and smaller parameters will remain secure~\cite{Grover2004collision,Bernstein2009Cost}. On the other hand, no lower bound establishes that comparable quantum speedups must incur such costs. Quantum time-space tradeoffs thus serve a practical role in analyzing the real-world security of cryptography~\cite{CNS2017}; for example, a more space-efficient quantum collision-finding algorithm could force a reassessment of hash output lengths.

\subsection{Compressed oracles and their limitations}\label{sec:compressedlimitations}
We now discuss the Compressed Oracle method of Zhandry~\cite{zhandry2019record} and its limitations for proving space lower bounds in general; in particular, we highlight the limitations for using the method to prove a time-space tradeoff for collision-finding. We discuss the applicability of other techniques and recent results in \cref{sec:lowerboundtechniques}.

\paragraph{Compressed Oracle background.} The Compressed Oracle method views the oracle register in a basis whose elements $D$ are unordered lists of input/output pairs $(x,y)$; $D$ is called a ``database.''  Intuitively, the records in $D$ represent the queries the adversary has ``seen.'' For example, if the adversary has just made a query on $x$ and received output $y$, then most of the amplitude will be on $D$ containing $(x,y)$. If the adversary later queries on $x$ to uncompute $y$, then $(x,y)$ will be removed from $D$. This intuition isn't strictly true, as there are some small errors and other caveats. However, certain formal statements can be proved. For example, after $t$ queries, the oracle register will be supported only on $D$ containing at most $t$ entries. Additionally, up to small error, an adversary can only ever output collisions which are also contained in $D$ at the same time.

\paragraph{Compressed Oracles versus collision-finding.} Zhandry reproves the space-agnostic query lower bound using compressed oracles by showing that, in each query, $\sqrt{p}$ can only increase by $\Os(\sqrt{T/N})$, where $p$ is the probability that $D$ contains a collision. Over $T$ queries, this gives $p=\Os(T^3/N)$. Achieving a constant success probability $p = \Omega(1)$ then gives the lower bound on $T$.

Inspecting Zhandry's proof, we can actually immediately bound the per-query progress by $\Os(\sqrt{|D|/N})$, where $|D|$ is the number of records in the database $D$.\footnote{Technically, the oracle register can be in superposition over different-sized databases, and the actual upper bound would replace $|D|$ with the expected collision-free database size.} A natural but incorrect attempt to prove a time-space lower bound is the following. If $D$ represented what the adversary ``knows'' about the oracle, and the adversary is space-constrained, then we might expect $|D|\log N$ to be bounded by the adversary's storage $S$. The $\log N$ term comes from the bit-length of the $y$ value in each record in $D$.\footnote{The actual bit-length would be $\log(MN)$ to also include the $x$ values. But the $x$ values might be fixed as $0,1,2,\cdots$ or some other predictable pattern, and wouldn't need to be stored by the adversary. However, the $y$ values are supposed to represent random oracle outputs, so we might expect we need to pay the full price of $\log N$ to store each one.} If such a bound on $|D|\log N$ could be proved, this would immediately prove that the per-query progress is $\Os(\sqrt{S/(N\log N)})$, yielding the desired $T^2 S=\Omega(N\log N)$ bound matching BHT. 

However, the intuitive bound of $|D|\log N = \Os(S)$ turns out to be false in general, as the next two examples show.

\paragraph{Database size is unrelated to adversary space.} We give a simplified view of the compressed oracle evolution in two scenarios. First, consider an adversary that simply computes $V=\Hm(1)\oplus \Hm(2)\oplus\cdots\oplus \Hm(T)$. $V$ can be computed in very small space $\Os(\log N)$ by simply keeping a running sum $(\bmod\ 2)$ of the various $\Hm$ outputs. However, if we look at the joint state of the adversary and compressed oracle, it will (roughly) correspond to
\begin{equation}
    \sum_{V}\ket{V}\otimes \sum_{\substack{y_1,\ldots,y_T\\y_1\oplus y_2\oplus\cdots\oplus y_T=V}} \ket{\;\{(1,y_1),(2,y_2),\cdots,(T,y_T)\}\;},
\end{equation}
ignoring normalization. This makes intuitive sense: the low-space algorithm doesn't ``know'' any particular $\Hm(i)$ value, since it is only ever storing the partial sums. However, it is storing a value that crucially depends on each $\Hm(i)$. This means each $(i,\Hm(i))$ input/output pair must be present in the database as $(i,y_i)$, but they will be completely uniform and independent of the adversary's view except for the overall XOR constraint. Thus, we see that the compressed oracle database can be much larger than the adversary's actual storage.

For a more natural and pertinent example, we consider the classical Pollard's rho algorithm implemented on a quantum computer. Assume for simplicity that $M=N$ (both here and for the remainder of the paper; our results easily extend to the $M>N$ case however). In Pollard's rho algorithm, we compute $\Hm^{(k)}(x)\triangleq\Hm(\Hm(\cdots \Hm(x)\cdots ))$, where $\Hm$ is applied a total of $k$ times. Imagine computing $\Hm^{(k)}(x)$ on a quantum computer, and uncomputing all the intermediate values. The end result is that the joint state of the adversary and oracle roughly follows the following schematic representation, with amplitudes and detailed oracle compression terms suppressed:
\begin{equation}
    \sum_{x,z}\ket{x,z}\otimes \sum_{y_1,\ldots,y_{k-1}}\ket{\;\{(x,y_1),(y_1,y_2),\cdots,(y_{k-2},y_{k-1}),(y_{k-1},z)\}\;}.
\end{equation}
Again, this intuitively makes sense: after the adversary computes the entire chain, but before uncomputing, it will ``know'' the input/output pairs $(\Hm^{(i)}(x),\Hm^{(i+1)}(x))$, so they will be present in the database as $(y_i,y_{i+1})$. Then when the adversary uncomputes the intermediate values, it does not remove the entries from the database, but just makes them independent of the adversary's view. Only the initial point $x$ and the final point $z$ will remain in the adversary's register.

If we set $k=\sqrt{N}$, then the standard birthday paradox argument shows that the chain of evaluations $x\rightarrow \Hm^{(1)}(x)\rightarrow \cdots\rightarrow \Hm^{(k)}(x)=z$ is likely to contain a collision; this manifests in the Compressed Oracle method by yielding a database $D$ that contains a collision. However, the adversary is unlikely to actually know the colliding points, since it only knows the first and last values $(x,z)$. Hence, while a collision exists, additional work is needed to actually extract the collision and output it.

The two examples above demonstrate that, while the compressed oracle in a sense \emph{upper-bounds} the adversary's knowledge of a random oracle, it is not tight; the compressed oracle can actually contain significant information that is unknown to the adversary. This leads us to ask:

\begin{center}
    \textbf{\emph{Can the Compressed Oracle method be refined \\to better capture what the adversary actually knows?}}
\end{center}

\paragraph{Label symmetry.} Very recently, Magniez and Zur~\cite{magniez2026label} prove an optimal $T^2S=\Omega(N\log N)$ tradeoff for quantum collision-finding, but only for algorithms that are ``label-symmetric.'' Label symmetry roughly means that the algorithm is promised to ``behave identically'', even if the outputs of the oracle are re-labeled by an arbitrary permutation. This captures many natural algorithms such as Grover search on a stored oracle label, the standard classical birthday attack, and BHT, as well as their straightforward space-limited generalizations. By giving conditions under which a provable lower bound holds, their work offers a promising new direction for studying time-space tradeoffs.

Very roughly, their lower bound technique starts with the Compressed Oracle method. However, they then prove directly that, for label-symmetric algorithms, the expected database size of collision-free databases is at most $2S/\log N$; this implies each query can only increase $\sqrt{p}$ by at most $\Os(\sqrt{S/(N\log N)})$, which is sufficient to plug into the outline above to obtain the desired $T^2S=\Omega(N\log N)$ bound. Rather than actually use the space $S$ in the bound,~\cite{magniez2026label} utilizes the Schmidt rank, which is upper-bounded by $2^S$. The analysis of the database size bound utilizes sophisticated techniques involving the representation theory of the symmetric group. Roughly, label symmetry is used to lower bound the Schmidt rank on algorithms with large amplitude on large collision-free databases. Magniez and Zur leave as an open problem removing the label symmetry restriction to get an unrestricted lower bound. To understand the feasibility of this goal, we ask:

\begin{center}
    \textbf{\emph{What is the role of label symmetry in lower-bounding the adversary's space?}}
\end{center}

\subsection{Our work}
We are now ready to discuss our results in detail. 

\paragraph{Dual compressed oracles.} First, motivated by the gap between recorded and recoverable information in the Compressed Oracle method, we develop a new technique to give a tighter understanding of the knowledge of quantum adversaries that do not make intermediate measurements.\footnote{Our framework is primarily defined in this setting, and it is the one mainly considered by Magniez--Zur~\cite{magniez2026label}. Recently, an updated version of~\cite{magniez2026label} on arXiv was added which considers a certain notion of label symmetry under intermediate measurements. We are able to extend our own simplified proof to handle this setting as well (see \cref{cor:intermediate-measurements}).} The central observation is that, if we have a joint adversary-oracle system $\sum_{z,D}\alpha_{z,D}\ket{z}\otimes \ket{D}$, we can apply a (partial-)isometry to view the adversary's register in a basis such that the adversary's state is dual to the oracle. This means that the state is
\[\sum_{\tilde{D},D}\beta_{\tilde{D},D}\ket{\tilde{D}}\otimes \ket{D}\qquad\text{where}\qquad \left(\beta\right)_{\tilde{D},D}\ \text{is positive semi-definite (PSD)}.\]
This view has multiple interpretations. It can be seen as performing a polar decomposition on the matrix $\left(\alpha_{z,D}\right)_{z,D}$. The matrix $\left(\beta\right)_{\tilde{D},D}$ is also the (PSD) square-root of the Gram matrix $G_{\tilde{D},D}=\sum_z \alpha_{z,\tilde{D}}^*\alpha_{z,D}=\sqrt{p_{D}p_{\tilde{D}}}\braket{\psi^{\tilde{D}}|\psi^D}$, where $\sqrt{p_D}\ket{\psi^D}$ is the (unnormalized) state of the adversary when conditioned on compressed oracle state $\ket{D}$. Note that, while there may be multiple choices for the basis, the resulting state is canonical; we call it the \emph{canonical representation} of the joint adversary-oracle system.

If we just examine the marginals $D$ or $\tilde{D}$ independently, they look exactly as in the standard Compressed Oracle method. However, the correlations between $D$ and $\tilde{D}$ reveal more fine-grained information. For example, consider an adversary which outputs a guess $(x,y)$ and wins if $y=\Hm(x)$. Viewing the canonical representation of the system and measuring $D$ and $\tilde{D}$, we show that the adversary needs to have $D$ and $\tilde{D}$ contain a common input/output pair in order to win with high probability. This qualitative statement is also tight in some sense; if $\tilde{D}$ and $D$ overlap, then there is an adversary which can win with reasonable probability without making any extra queries (note, however, that the quantitative upper and lower bounds have a slack, see \cref{subsec:info-extraction}). Similarly, in order to output a collision with high probability, it must be that $D$ and $\tilde{D}$ contain the same collision.

Note that in the XOR case above, the canonical representation roughly results in
\begin{equation}
    \sum_{\substack{y_1,\ldots,y_T\\\tilde{y}_1,\ldots,\tilde{y}_T\\y_1\oplus\cdots\oplus y_T=\tilde{y}_1\oplus\cdots\oplus\tilde{y}_T}}\ket{\{(1,\tilde{y}_1),(2,\tilde{y}_2),\cdots,(T,\tilde{y}_T)\}}\otimes \ket{\{(1,y_1),(2,y_2),\cdots,(T,y_T)\}}
\end{equation}
We see that, with high probability, $y_i\neq \tilde{y}_i$ for all $i$, despite the global constraint $y_1\oplus\cdots\oplus y_T=\tilde{y}_1\oplus\cdots\oplus\tilde{y}_T$. Applying our results from Compressed Oracle Duality, the adversary is unable to output any input/output pair. This matches our understanding developed above.

Likewise, in the Pollard rho case, the canonical representation has the following schematic form:
\begin{equation}
    \sum_{x,z}\sum_{\substack{y_1,\ldots,y_{k-1}\\\tilde{y}_1,\ldots,\tilde{y}_{k-1}}}\ket{\{(x,\tilde{y}_1),(\tilde{y}_1,\tilde{y}_2),\ldots,(\tilde{y}_{k-1},z)\}}\otimes\ket{\{(x,y_1),(y_1,y_2),\ldots,(y_{k-1},z)\}}
\end{equation}
By the birthday bound, if $k=\sqrt{N}$, both sides are likely to have a collision. However, it is very unlikely that the collisions on either side agree. Hence, by the output probability upper bound, the adversary is unable to actually output a collision at this point. The adversary needs to perform extra work in order to force $\tilde{D}$ and $D$ to have a matching collision, at which point it may output a collision.\footnote{Additionally, this example can be made to use low space at all points in time, by pebbling the chain instead~\cite{bennett1989time,Banegas2017search,jaques2020lowgate}.}

\paragraph{A simple proof of the label-symmetric lower bound.} We use our new framework to give a simpler proof of the time-space lower bound for label-symmetric collision-finding algorithms. Crucially, by using Compressed Oracle Duality instead of the standard Compressed Oracle method, we are able to avoid the complicated representation theory required by~\cite{magniez2026label}. Our simplified proof outline is the following:
\begin{itemize}
    \item Following existing analysis, it suffices to bound the progress on $\sqrt{p}$ by bounding the database size on the portion of the state where $D$ contains no collisions $\sum_{D: \CF(D)}G_{D,D}|D|$. $\CF(D)$ denotes that $D$ does not contain any collisions, and we informally call this summation the ``collision-free database size'' here.
    \item Let the ``label set'' of a database $D$ be the set of $y$-values; use $Y$ to denote the label set of $D$ and $\tilde{Y}$ to denote the label set of $\tilde{D}$. We prove that, for query-bounded label-symmetric algorithms, the expected contribution of the ``unmatched'' label set $|Y\setminus \tilde{Y}|$ after $t$ queries to the collision-free database size is roughly $t^2/N$. This is shown using label symmetry and validity (compressed oracle entries are orthogonal to $\ket{\hat{0}}$), which implies the amplitude on unmatched labels is uniform among all labels not in either database and therefore must be small in magnitude. 
    \item Next, we show that the size of the ``matched'' label set $|Y\cap \tilde{Y}|$ (intuitively, the labels that the adversary ``has knowledge of'') roughly lower bounds the space of the adversary $S$. The number of possible label sets for $Y\cap \tilde{Y}$ is $\binom{N}{|Y\cap \tilde{Y}|}\approx \frac{N^{|Y\cap \tilde{Y}|}}{|Y\cap \tilde{Y}|!}$.\footnote{Again, the inputs $x$ could all be fixed. Only the labels necessarily contribute to the number of databases.} Since these labels are essentially uniform under label symmetry, matched labels between $Y$ and $\tilde{Y}$ give a mutual information which is roughly lower-bounded by the logarithm of the number of matched-label sets. This mutual information is also bounded by $\log \rank(G)\leq S$ due to the Holevo bound, which gives a contribution of $|Y\cap \tilde{Y}|\lesssim S/\log N$ on collision-free database size.
    
    \item Combining $|Y\cap \tilde{Y}|$ and $|Y\setminus \tilde{Y}|$ gives $|Y|=|D|$ on collision-free databases, so we get the upper bound on collision-free database size of $\Os(S/\log N)$, allowing us to finish the earlier failed intuitive proof strategy.
\end{itemize}

To deal with the intermediate measurements, we ``label-symmetrize'' each conditional branch state by applying a twirling operator over the symmetric group. This preserves the rank upper bound of the branch Gram matrix while also simultaneously making it satisfy label symmetry, allowing the original proof to apply.

\paragraph{A better understanding of label symmetry.} While the time-space lower bound for label-symmetric algorithms was already proved by~\cite{magniez2026label}, our proof yields a better understanding of the role that label symmetry plays in the lower bound.

Recall the example of Pollard's rho, which is used to find collisions classically in low space. Pollard's rho is \emph{not} label-symmetric. Indeed, in the chain $x\rightarrow y_1=\Hm^{(1)}(x)\rightarrow y_2=\Hm^{(2)}(x)\rightarrow\cdots$, re-labeling $y_1$ to $x$ would result in the entire chain having the same value, which violates label symmetry. As such, the label-symmetric lower bound does not apply to algorithms utilizing similar techniques.

One may optimistically hope that similar ideas can be leveraged to give a full lower bound for general algorithms including Pollard's rho, removing the label symmetry restriction. However, our new proof indicates that this is unlikely without fundamentally new ideas. Indeed, the Pollard's rho example shows that the adversary's space is entirely disconnected from the size of $D$. On the other hand, the lower bound arising from label symmetry works exactly because it forces the collision-free database size to be essentially $\Os(S/\log N)$, including when intermediate measurements are made!

To add to this evidence (while also showcasing the utility of our new techniques), we demonstrate that among \emph{classical} algorithms, label symmetry implies a $TS=\Omega(N\log N)$ time-space lower bound. However, the Pollard's rho algorithm achieves $TS=\Os(N^{1/2}\log N)$, showing that label symmetry actually gives the \emph{incorrect} time-space tradeoff for classical algorithms. Our proof defines the matched-label set of a classical memory state, which is the classical analogue of the labels shared by $D$ and $\tilde{D}$. The $2S/\log N$ bound on its size (\cref{lem:classical-support-space}) follows from the same Compressed Oracle Duality and Holevo argument as the quantum proof. Strengthening the tradeoff scaling from $T^2S$ to $TS$ additionally requires comparing the branching program's execution on $\Hm$ with its execution on a collision-free injective oracle (\cref{lem:classical-label-progress}).

\subsection{Future outlook on quantum time-space tradeoffs}
The decoupling between adversary space and the size of the compressed oracle database suggests that bounding database size by itself is an insufficient path towards proving an unrestricted quantum time-space tradeoff for collision-finding. In the Pollard's rho compressed oracle example, we see that a database can be very large but remain inaccessible to the adversary without making additional oracle queries.

\paragraph{A proposal for collision-finding using Compressed Oracle Duality.} A natural approach to show a collision-finding time-space tradeoff using our framework is the following: instead of using database size as a bound on the per-query progress towards finding a collision, one should instead (1) bound the number of points \emph{the adversary can actually access} in the compressed oracle database, then (2) show that collision-finding progress is bounded by the number of accessible points in the same manner that database size bounds progress.

We formalize the concept of ``accessible information'' in \cref{subsec:info-extraction} by defining a natural class of potential functions which count the number of points which appear in \emph{both the compressed oracle and its dual simultaneously}. Thus we can define both a \emph{joint collision potential} $\Phi_{\text{collision}}$ which bounds the probability of finding a collision and a \emph{joint point potential} $\Phi_{\text{point}}$ which bounds the total number of database points an adversary can output at any time. Furthermore, we see in \cref{subsec:candidate-approach} that adversary space does indeed upper bound the number of accessible points by showing $\Phi_{\text{point}} = \Os(S/\log N)$. The final step of showing 
\begin{equation}
    \Phi_{\text{collision}}(t+1)-\Phi_{\text{collision}}(t) = \Os(\sqrt{\Phi_{\text{point}}/N})
\end{equation}
would then imply the full unrestricted tradeoff. Unfortunately it can be shown that, as presently defined, there exist counterexamples refuting this bound.\footnote{We briefly sketch one (contrived) example: an adversary which first finds a collision $(x_1, x_2, y)$, say using Pollard's rho, then ``hides it'' by XORing $y$ with an arbitrary oracle query to get $\Hm(x')\oplus y$. Then the adversary reveals $y$ again by once again XORing with the same query: $\Hm(x')\oplus (\Hm(x')\oplus y) = y$. It can be shown that this final query results in a very large increase in the joint potential independent of adversary space.}

However, since Compressed Oracle Duality distinguishes between information which exists in a compressed oracle and the information that can be accessed by an adversary, we hope this method can be used to prove an improved unrestricted tradeoff (for quantum algorithms without intermediate measurements); for example, we conjecture that a suitable modification of the joint potential definitions in the pair $(\Phi_{\text{collision}}, \Phi_{\text{point}})$ may allow for the desired per-query upper bound.

\paragraph{The restriction to unitary algorithms.}
As currently presented, Compressed Oracle Duality applies only to quantum algorithms without intermediate measurements. The standard Compressed Oracle method does not need this restriction since for query lower bounds measurements can be deferred to the end of the algorithm at no cost without changing the output distribution. For time-space tradeoffs, deferring a measurement requires an additional qubit to hold its outcome, whereas a space-bounded adversary can measure qubits and reuse them (there exist some results which allow for deferred measurements, but they generally still come at a cost to the time-space tradeoff~\cite{GR22,Zha24}).

Conditioning on the history of measurement outcomes keeps each branch in a pure state with memory of dimension at most $2^S$, so the canonical representation remains well defined branch by branch. However, the canonical representation no longer captures what the adversary knows: an outcome that the adversary measures and then discards still determines the compressed oracle's conditional state, so it also appears in the dual compressed oracle. For example, if the adversary measures $\Hm(x)=y$ and resets the register, then in that branch both $\tilde{\Dm}$ and $\Dm$ contain $(x,y)$, although the memory holds nothing about $y$.\footnote{The label symmetry condition of Magniez--Zur essentially forces the adversary to store $y$ in memory, further suggesting that the assumption may be overly strong and potentially exclude space-efficient algorithms.}
Providing a method that captures the adversary's knowledge after intermediate measurements remains an interesting open research direction.

\subsection{On existing lower bound techniques} \label{sec:lowerboundtechniques}

For completeness, we survey the currently known lower bound techniques in query complexity and explain why using them to obtain space lower bounds is so difficult.

\paragraph{The Hybrid method.} The hybrid method of~\cite{Bennett_1997} bounds how much the state of the adversary changes in Euclidean distance as points in the oracle are changed. The dimension of the state (which corresponds to space) is irrelevant. 

\paragraph{The Polynomial method.} The polynomial method of~\cite{beals01} shows that the probability of any particular output of a $T$-query quantum algorithm is a degree-$2T$ polynomial. Additionally, the polynomial, being a probability, must attain values in $[0,1]$ on all oracles. This method then analyzes the possibility of such low-degree bounded polynomials, completely ignoring that the polynomial came from a quantum algorithm. In particular, the space $S$ has no influence on the degree and hence the query bounds. 

\paragraph{The Adversary method.} The (negative weights) adversary method~\cite{ambainis2000quantum,BSS_2003,HLS07}, which generalizes the hybrid method, converts the task of proving a query complexity lower bound into solving a derived semi-definite program (SDP). This is a powerful technique, in the sense that the fully general negative weights method can prove optimal strictly-time-based query lower bounds for any function~\cite{reichardt2011span,Lee2011Quantum}. However, the method does not directly relate to the adversary's space, and the characterization of query complexity as SDPs is space-agnostic.

Additionally, the characterization of query complexity as SDPs, even if it could be generalized to also consider space, does not immediately yield a way to solve the query complexity question, and often the obtained SDP seems just as hard to analyze as the original question. In a sense, the negative weights adversary method is too powerful: since it captures \emph{everything}, solving the SDP in general is necessarily as hard as solving arbitrary query complexity problems. For this reason, most new query complexity lower bounds are actually proved via restricted versions of the adversary method or other methods entirely, as they are often easier to use. 

\paragraph{The Compressed Oracle method.} The Compressed Oracle method has enabled many new query lower bounds. However, as discussed above, this method also does not directly connect to space.

\paragraph{Existing quantum time-space tradeoffs.} Despite the aforementioned difficulties, some quantum time-space tradeoffs have been shown, such as for large-output problems~\cite{klauck2007quantum,hamoudi2023quantum,beam2026matrix}, or in the setting of size-bounded advice (but ignoring ``online'' space)~\cite{chung2020function,liu2023nonuniformity}. Typically, these results work by employing standard space-agnostic quantum query lower bound techniques but then separately folding in a space constraint. For example, \cite{hamoudi2023quantum} and~\cite{chung2020function} incorporate space by actually completely discarding the adversary's space at a particular point in time and then guessing the state from scratch. For an $S$-qubit/bit algorithm, this guessing step incurs a loss factor of $2^S$; then standard query complexity techniques are used to prove a strong success probability bound of $2^{-s(T)}$. A successful adversary must therefore have $S\geq s(T)$. 

The limitation of advice-guessing is that the success probability needs to be an extremely small exponential in order to absorb the exponential advice-guessing loss. This makes it suitable for certain large-output problems that have such tiny success probabilities. But for problems with small outputs of length $\ell$, the success probability is always at least $2^{-\ell}$ by random guessing. Hence, for such problems, simple advice guessing seems limited to space smaller than $\ell$. This precludes obvious application to collision-finding.

This ``advice guessing'' technique combined with compressed oracles was utilized in the online, small-output setting in the recent breakthrough of Hao, Huang, and Liu \cite{Hao2026short}, who showed a time-space tradeoff for a particular ``nested-collision'' problem. At a very high level, the proof involves showing that any algorithm making progress on computing a nested collision must actually be implicitly solving a large-output problem, in the sense that a solution to the large-output problem must be present in the compressed oracle database. The large-output problem is essentially to find $K$ distinct tuples $(x_1,\cdots,x_\ell)$ satisfying $H(x_1)+H(x_2)+\cdots +H(x_\ell)=y\bmod N$, for some specified target $y$.

Currently, \cite{Hao2026short,gao2026query} are the only known online, small-output quantum time-space tradeoffs that make no restrictions on the algorithm's behavior aside from bounding its time and space. Unfortunately, the nested-collision problem of \cite{Hao2026short} and the filtered-parity problem of \cite{gao2026query} are somewhat contrived, and their techniques do not seem directly useful for the standard collision problem. In particular, there is no obvious large-output problem that needs to be solved in order to make progress on finding a single collision.

\subsection*{Acknowledgments}
M.D. is supported in part by the National Science Foundation Graduate Research Fellowship Program under Grant No.\ DGE-2146755. M.Z. is supported by an Amazon Research Award.

\subsection*{AI disclosure}
The key ideas behind the Compressed Oracle Duality were developed solely by the authors, independent of AI and the work of Magniez--Zur \cite{magniez2026label}. After the release of \cite{magniez2026label}, the authors used ChatGPT 6 Astra along with this new framework to develop the simplified proofs. The authors contributed all writing and exposition, while ChatGPT and Claude assisted with proofreading and minor suggestions. The authors take responsibility for the originality and correctness of all proofs.

\input{preliminaries}
\input{duality-framework}
\input{quantum-label}
\input{classical-label}

\printbibliography

\end{document}

%% file: preliminaries.tex
\section{Preliminaries and notation}
Random variables are denoted in bold. All logarithms are base two. $\Hm:[N]\to[N]$ denotes a uniformly random oracle, where $N \triangleq 2^n$ and $[N]\cong \{0,1\}^n$. $S$ denotes the space bound of an algorithm. For quantum algorithms, the space includes all qubits and classical bits. We assume $S\geq \log N$ and $N\geq 64$ throughout. For a set $X\subseteq [N]$, let $\Hm|_{X}$ denote the restriction of the random oracle outputs of $\Hm$ to domain $X$. For a set $X$, $\binom{X}{k}$ denotes the set of all size-$k$ subsets of $X$.

\paragraph{Vectorization.}
Fix orthonormal bases $\{\ket a\}_{a\in A}$ and $\{\ket b\}_{b\in B}$ of the finite-dimensional Hilbert spaces $\Hs_A$ and $\Hs_B$. We use the linear bijection
\begin{equation}
    \mathsf{vec}:\Ls(\Hs_B,\Hs_A)\longrightarrow\Hs_A\otimes\Hs_B,
    \qquad
    \mathsf{vec}(\ket a\bra b)=\ket a\otimes\ket b.
\end{equation}
The following standard identity is given in \cite[Eq.~(1.132)]{Watrous_2018}.

\begin{proposition}\label{prop:vectorization-pullback}
    Let $\Hs_A,\Hs_B,\Hs_C,\Hs_D$ be finite-dimensional
    Hilbert spaces with specified orthonormal bases, and let
    \begin{equation}
        X\in\Ls(\Hs_B,\Hs_A),\qquad
        Y\in\Ls(\Hs_C,\Hs_B),\qquad
        Z\in\Ls(\Hs_D,\Hs_C).
    \end{equation}
    Then
    \begin{equation}
        \mathsf{vec}(XYZ) =(X\otimes Z^\top)\mathsf{vec}(Y).
    \end{equation}
\end{proposition}

\subsection{Information theory}
The following results are all from \cite[Chapter~5]{Watrous_2018}. $\mathrm{H}(\Xm)$ denotes the entropy of random variable $\Xm$ measured in bits. $I(\Xm;\Ym)\triangleq \mathrm{H}(\Xm)-\mathrm{H}(\Xm\mid \Ym)$ denotes the mutual information. We have the following facts:
\begin{enumerate}
    \item Entropy chain-rule: 
    \begin{equation}
        \mathrm{H}(\Xm,\Ym) =  \mathrm{H}(\Xm) + \mathrm{H}(\Ym\mid \Xm).
    \end{equation}

    \item Log-sum inequality: For $a_i\geq 0, b_i > 0$ for $i\in[n]$, $a\triangleq \sum_{i=1}^n a_i$, and $b\triangleq \sum_{i=1}^n b_i$,
    \begin{equation}
        \sum_{i=1}^n a_i \log\left(\frac{a_i}{b_i}\right) \geq a \log\left(\frac{a}{b}\right),
    \end{equation}
    where $0 \log 0 \triangleq 0$.
\end{enumerate}

To bound the information from a quantum measurement, we use a standard consequence of the Holevo bound:
\begin{lemma}\label{lem:holevo-bound}
    Let $\{(p_i,\ket{\psi_i})\}_i$ be a finite ensemble of pure states $\ket{\psi_i}$ with prior $p_i$ spanning a subspace of dimension $r$, and let $\{E_j\}_j$ be a POVM on that subspace. Let $\Xm$ be the ensemble index and $\Ym$ the measurement index, with joint distribution
    \begin{equation}
        \Pr[\Xm=i,\Ym=j]=p_i\braket{\psi_i|E_j|\psi_i}.
    \end{equation}
    Then
    \begin{equation}\label{eq:rank-test}
        I(\Xm;\Ym)\leq\log r.
    \end{equation}
\end{lemma}

\paragraph{Square-root measurement.}
Let $\{(p_i,\rho_i)\}_i$ be a finite ensemble of quantum states with density matrix $\rho=\sum_i p_i\rho_i$. The \emph{square-root measurement}, also called the
\emph{pretty-good measurement}~\cite{HW94,EldarForney2001}, is the POVM on the support of $\rho$ given by
\begin{equation}\label{eq:square-root-measurement}
    E_i\triangleq \rho^{-1/2}(p_i\rho_i)\rho^{-1/2},
\end{equation}
where the inverse square root is taken on the support of $\rho$ and is zero on its orthogonal complement.

\subsection{Quantum query complexity}
We first review the standard quantum query complexity formalism (e.g., \cite{hamoudi2025introduction}) for the setting our paper considers. Following the convention from cryptography, we also refer to the algorithm as the adversary, and use the terms interchangeably.

The adversary uses a total of $S$ qubits of space, and we denote the adversary Hilbert space $\Hs_{\mathsf{A}}\cong \C^{2^S}$. This space decomposes as the tensor product of the input register $\Hs_{\mathsf{X}}\cong \C^{N}$, output register $\Hs_{\mathsf{Y}}\cong \C^{N}$, and the remaining workspace register $\Hs_{\mathsf{W}}$. Access to input $H$ is through an oracle unitary $\mathsf{O}_H\in\Ls(\Hs_{\mathsf{A}},\Hs_{\mathsf{A}})$, which acts on the query registers as
\begin{equation}
    \mathsf{O}_H\ket{x}_{\mathsf{X}}\ket{y}_{\mathsf{Y}}\ket{w}_{\mathsf{W}} = \ket{x}_{\mathsf{X}}\ket{y\oplus H(x)}_{\mathsf{Y}}\ket{w}_{\mathsf{W}},\qquad x,y\in [N], w\in\{0,1\}^{S-2n}.
\end{equation}

We consider the canonical form of unitary quantum query algorithms, i.e., quantum query algorithms with no intermediate measurements on the adversary registers. Our framework is defined in this setting, but we give an extension specifically for the label-symmetric proof in \cref{subsec:intermediate-measurements}.
\begin{definition}[Canonical form of unitary quantum query algorithms]
    For a quantum query algorithm $\Alg$ using $T$ queries to input $H$ and $S$ qubits of space which applies oracle-independent unitary $V_t$ at time step $t$, we define the intermediate state of the adversary at time $t \in \{0, 1, \ldots, T\}$ as
    \begin{equation}
        \ket{\psi_t^H} \triangleq V_t\mathsf{O}_H V_{t-1}\mathsf{O}_H\cdots V_1 \mathsf{O}_H V_0\ket{0^S}.
    \end{equation}
\end{definition}

In the (Quantum) Random Oracle model~\cite{BR93,Boneh_2011}, the input is a uniformly sampled random oracle $H\sim \mathbf{H}$. The state of the adversary is a mixed state, and it is generally convenient to work with a purification in which an additional register records the input. Consider an $S$-qubit adversary mixed state with density matrix $\rho = \sum_{H}p_H\ket{\psi^H}\bra{\psi^H}$ over an ensemble of inputs $\{H\}_{H}$. Its purification is represented as $\ket{\psi} \in \Hs_{\mathsf{A}} \otimes \Hs_{\mathsf{H}}$, given by
\begin{equation}
    \ket{\psi} = \sum_H \sqrt{p_H}\ket{\psi^H}_{\mathsf{A}} \otimes \ket{H}_{\mathsf{H}},
\end{equation} 
where $\Hs_{\mathsf{H}}\cong \bigotimes_{x\in [N]} \C^{N}$ is spanned by the basis states $\{\ket{H}\}_{H}$. We call $\psi$ the \emph{joint adversary-oracle state}.

\subsection{Compressed Oracle method}
We utilize the Compressed Oracle method of Zhandry~\cite{zhandry2019record}. Let $\Hs_{\mathsf{D}} = \bigotimes_{x\in [N]} \C^{N + 1}$ denote the Hilbert space of the compressed oracle, where each local compressed oracle register (or ``cell'') is indexed and spanned by output basis states $\{\ket{y}\}_{y\in [N]} \cup \{\ket{\bot}\}$. We also use the term ``computational database basis/state'', which refers to a state $\ket{D}$, where $D$ refers to a single oracle database. The set $\{\ket{D}\}_{D}$ forms a canonical basis over Hilbert space $\Hs_{\mathsf{D}}$. The compressed oracle is initialized to the empty database $\ket{\bot}^{\otimes N}$. We use $D[x]$ to denote the specific value of $D$ given at cell $x$; $x$ is called the \emph{input} and $D[x]$ is called the \emph{label}, and $D[x]=\bot$ signifies that no information about $H(x)$ has been recorded. Denote $\supp(D)\triangleq\{x:D[x]\neq\bot\}$, $|D|\triangleq |\supp(D)|$, and $\range(D)\triangleq\{D[x]:x\in\supp(D)\}$. Let $\CF(D)$ denote that $D$ is a collision-free database, i.e. $D$ restricted to domain $\supp(D)$ is injective.

Let $\Pi_\mathsf{X}(x)$ and $\Pi_\mathsf{Y}(y)$ denote the projectors onto basis states $\ket{x}$ in the input register and $\ket{y}$ in the output register respectively. $\ket{\hat{0}} \triangleq N^{-1/2}\sum_{y\in[N]}\ket{y}$ denotes the uniform superposition over outputs, and $\StdDecomp_x$ denotes the unitary acting on the register $x$ of $\mathsf{D}$ which swaps $\ket{\bot}$ and $\ket{\hat{0}}$ and acts as the identity on the orthogonal complement of $\operatorname{span}\{\ket{\hat{0}},\ket{\bot}\}$; note that $\StdDecomp_x$ is Hermitian, so that $\StdDecomp_x^2=\mathbb{I}$. 

We can write the action of the standard compressed oracle query $\CStO: \Hs_{\mathsf{A}} \otimes \Hs_{\mathsf{D}} \rightarrow \Hs_{\mathsf{A}} \otimes \Hs_{\mathsf{D}}$ as
\begin{equation}
    \sum_{x \in [N]} \Pi_\mathsf{X}(x) \left(\StdDecomp_x \circ \StO_x \circ \StdDecomp_x\right)
\end{equation}
where $\StO_x$ XORs the value of $D[x]$ into $\mathsf{Y}$ and acts as the identity when $D[x]=\bot$. The following is the key lemma which relates the Compressed Oracle method to the standard query model over a random oracle $H$.

\begin{lemma}[Two-sided form of {\cite[Lemma~5]{zhandry2019record}}]\label{lemma:compressed_oracle}
    Let $K\subseteq [N]^\ell \times [N]^\ell$ be a set of $\ell$-tuples of $(x,y)$ pairs. Consider a quantum query algorithm $\Alg$ in the Quantum Random Oracle model that either outputs vectors $(\Vec{x}, \Vec{y}) \in [N]^\ell \times [N]^\ell$, or $\bot$. Let $p$ denote the probability that $\Alg$ outputs $(\Vec{x}, \Vec{y})\in K$ such that $\Hm(x_i)=y_i$ for all $i\in [\ell]$. Let $p'$ denote the probability that $\Alg$ simulated with a compressed oracle outputs $(\Vec{x}, \Vec{y})\in K$ such that $\Dm[x_i]=y_i$ for all $i\in [\ell]$, where $\Dm$ is the measurement outcome of the compressed oracle database in register $\mathsf{D}$ at the end of the execution. Then
    \begin{equation}
        \left|\sqrt{p}-\sqrt{p'}\right|\leq\sqrt{\frac{\ell}{N}}.
    \end{equation}
\end{lemma}
\begin{proof}
    The bound follows from the operator-norm estimate in the
    proof of \cite[Lemma~4.1 and Corollary~4.2]{chung2021compressed}.
\end{proof}

Define the \emph{valid subspace} of compressed oracles, which is the tensor product of the orthogonal complement of $\ket{\hat{0}}\triangleq N^{-1/2}\sum_y\ket y$ in each cell. The valid subspace is closed under complex conjugation, because $\ket{\hat{0}}$ is real.

\paragraph{Database Gram matrix.} Write the joint adversary-oracle state (with a compressed oracle) as $\ket{\psi}=\sum_D\sqrt{p_D}\ket{\psi^D}_{\mathsf{A}}\ket D_{\mathsf{D}}$, where $p_D$ is the probability of measuring $D$ and the conditional memory state $\ket{\psi^D}$ is normalized whenever $p_D>0$. Fix an arbitrary orthonormal basis $\{\ket{z}\}_{z \in \{0,1\}^S}$ spanning the adversary Hilbert space $\Hs_{\mathsf{A}}$, and let the \emph{coefficient matrix} be the unique linear map $A: \Hs_{\mathsf{D}} \rightarrow \Hs_{\mathsf{A}}$ satisfying 
\begin{equation}
     A\ket{D}_{\mathsf{D}} = \sqrt{p_D}\ket{\psi^D}_{\mathsf{A}},\quad \ket{\psi} = \sum_{z,D} (A)_{z,D} \ket{z}_{\mathsf{A}} \otimes \ket{D}_{\mathsf{D}}.
\end{equation}
The \emph{Gram matrix} is the positive-semidefinite operator $G \in \Ls(\Hs_{\mathsf{D}}, \Hs_{\mathsf{D}})$ defined as
\begin{equation}
    G \triangleq A^\dagger A,\quad G_{\tilde{D},D}=\sqrt{p_{\tilde{D}}p_D}\braket{\psi^{\tilde{D}}|\psi^{D}}.
\end{equation}
We have $\Tr(G)=1$ and $\rank(G) \leq 2^S$. $G$ is supported on databases with at most $t$ entries. Call an operator \emph{valid} if its image lies in the valid subspace of compressed oracles. The Gram matrix $G$ is valid because $\ker G=\ker A$, so its image is the image of $A^\dagger$, which is spanned by complex conjugates of valid compressed oracle states.

The diagonal terms $G_{D,D} = p_D$ represent the probability that a computational basis measurement of $\mathsf{D}$ gives database $D$. Call the distribution induced by $p_D$ the \emph{standard database distribution} of $\ket{\psi}$. Additionally, the adversary reduced density matrix can be expressed as $\rho = AA^\dagger$, and the oracle reduced density matrix is given by $G^\top$. We use the subscript $t$ to denote the values after query $t$, e.g., $G_t, \ket{\psi_t}, \ket{\Psi_t}$, etc.

\paragraph{Database relations.}
A \emph{certificate} is a finite set $c=\{(x_1,y_1),\ldots,(x_\ell,y_\ell)\}$ of input--output pairs with distinct inputs and no $\bot$ outputs. $|c|\triangleq \ell$ denotes the size of the certificate. Identifying $H$ with the set $\{(x,H(x)):x\in[N]\}$, we say an oracle $H$ contains certificate $c$, written $c\subseteq H$, if $H(x_i)=y_i$ for all $i\in[\ell]$; likewise, identifying $D$ with the set $\{(x,D[x]):x\in\supp(D)\}$, a database $D$ contains $c$ if $D[x_i]=y_i$ for all $i\in[\ell]$, i.e.\ $c\subseteq D$. 
A \emph{relation} $\Rs$ is a finite set of certificates fixed independently of the oracle.

%% file: duality-framework.tex
\section{Compressed Oracle Duality}\label{sec:duality-framework}
In this section, we introduce a general framework for analyzing space-bounded quantum query algorithms. We follow the Compressed Oracle method but introduce an additional step for analyzing quantum memory by constructing a dual compressed oracle which is entangled with the original compressed oracle. We call this general framework \emph{Compressed Oracle Duality}. This section and the framework as a whole are completely independent of label symmetry.

\cref{subsec:duality-framework} introduces the definitions within the framework. \cref{subsec:info-extraction} gives a class of natural potential functions which can be defined using the framework and shows that these potential functions characterize optimal information-extraction from the adversary memory. Finally, we end with a proposal in \cref{subsec:candidate-approach} for how to prove time-space tradeoffs using our framework and describe the difficulties which remain.

\subsection{Defining the framework}\label{subsec:duality-framework}
Given a joint adversary-oracle state $\ket{\psi}$ with coefficient matrix $A$, our approach begins by the polar decomposition $A=U\sqrt{G}$, where $\sqrt G$ is the unique positive-semidefinite square root of $G$ and $U$ is the unique partial isometry satisfying $\ker U=\ker A$. Using vectorization, we define the following representation of the joint adversary-oracle state:
\begin{definition}[Dual compressed oracles]
    Given a joint adversary-oracle state $\ket{\psi} \in \Hs_{\mathsf{A}} \otimes \Hs_{\mathsf{D}}$ with Gram matrix $G$, we define its \emph{canonical representation} $\ket{\Psi} \in \Hs_{\mathsf{D}} \otimes \Hs_{\mathsf{D}}$ as
    \begin{equation}
        \ket{\Psi} \triangleq \mathsf{vec}\left(\sqrt{G}\right) = \sum_{\tilde{D},D} (\sqrt{G})_{\tilde{D},D} \ket{\tilde{D}}\ket{D}.
    \end{equation}
    We still identify the subsystem on the right side of the tensor product as the standard compressed oracle, but we now also identify the subsystem on the left side of the tensor product as the \emph{dual compressed oracle}.
\end{definition}

The matrices $G,\sqrt{G}$ and the state $\ket{\Psi}$ are all invariant under oracle-independent unitaries applied to the adversary, hence the ``canonical'' descriptor. The original state can also be viewed as the vectorization of the coefficient matrix: $\ket{\psi} = \mathsf{vec}(A)$. Using the vectorization identity \cref{prop:vectorization-pullback}, applying $U^\dagger$ to the adversary subsystem $\Hs_{\mathsf{A}}$ maps state $\ket{\psi}$ to its canonical representation:
\begin{equation}
    ((U^\dagger)_{\mathsf{A}}\otimes \mathbb{I}_{\mathsf{D}})\ket{\psi} = ((U^\dagger)_{\mathsf{A}}\otimes \mathbb{I}_{\mathsf{D}})\, \mathsf{vec}(A) = \mathsf{vec}(U^\dagger A) = \ket{\Psi}.
\end{equation}

The map $U^\dagger$ is an isometry from the support of $\rho$ to the image of $G$. Thus the dual compressed oracle is a canonical re-encoding of the adversary's memory, and every measurement on the dual oracle register induces an equivalent POVM on the original memory. In particular, measuring the dual oracle register in the database basis gives a distribution over databases, whose outcome we denote by $\tilde{\Dm}$. 

\begin{definition}[Joint database distribution]
    Define the \emph{joint database distribution} of $\ket{\psi}$ as the joint distribution $p(\tilde{D},D)$ of the outcomes $(\tilde{\Dm},\Dm)$ resulting from measuring both databases in $\ket{\Psi}$:
    \begin{equation}
        p(\tilde{D}, D) \triangleq |(\sqrt{G})_{\tilde{D}, D}|^2.
    \end{equation}
\end{definition}

Since $\sqrt{G}$ is Hermitian, $p(\tilde{D}, D) = p(D, \tilde{D})$. $\sqrt{G_t}$ is also valid (its image lies in the valid subspace of compressed oracles) since it is Hermitian with the same image as $G$ and each of its rows is orthogonal in every cell to $\ket{\hat{0}}$. Both marginals of the joint database distribution are identical to the standard database distribution:
\begin{equation}
    \sum_{D} p(\tilde{D}, D) = \sum_{D}|(\sqrt{G})_{\tilde{D}, D}|^2 = \sum_{D}(\sqrt{G})_{\tilde{D}, D}(\sqrt{G})_{D, \tilde{D}} = G_{\tilde{D}, \tilde{D}} = p(\tilde{D}).
\end{equation}
Consequently, after $t$ queries, both the compressed oracle and its dual lie in the valid subspace and measurement outcomes are supported on databases with size at most $t$.

\paragraph{A pretty-good interpretation.}
    Let $\rho\triangleq\sum_{D} p_D \ket{\psi^D}\bra{\psi^D}$ be the density matrix of the adversary subsystem. Measuring the dual oracle register in the database basis induces the following POVM on $\Hs_{\mathsf{A}}$:
    \begin{equation}
        \{E_{D}\}_{D} \, \cup \{E_\bot\},\qquad E_{D}
        \triangleq U\ket{D}\bra{D}U^\dagger
        =\rho^{-1/2}A\ket{D}\bra{D}
          A^\dagger\rho^{-1/2},\quad E_{\bot} \triangleq \mathbb{I}_{\mathsf{A}}-UU^\dagger,
\end{equation}
where the inverse is taken on the support of $\rho$. This POVM is exactly the square-root measurement for the ensemble $\{(p_D,\ket{\psi^D})\}_D$. Let $\Dm$ denote the ensemble label outcome and $\tilde{\Dm}$ the POVM outcome. Since $U^\dagger A=\sqrt{G}$, their joint distribution is
\begin{equation}
    \Pr[\tilde{\Dm}=\tilde{D}, \Dm=D]
    =\bra D A^\dagger E_{\tilde D}A\ket D
    =|(\sqrt G)_{\tilde D,D}|^2
    =p(\tilde D,D).
\end{equation}
Thus the joint database distribution is exactly the joint distribution of the database ensemble label and the outcome of the pretty-good measurement on the adversary memory.
    
The square-root measurement is also in some sense a near-optimal ``canonical measurement'' \cite{montanaro2007pgm}; in \cref{subsec:info-extraction}, we give analogous results for extracting information from quantum memory about a random oracle. Additionally, this interpretation combined with \cref{lem:holevo-bound} gives
\begin{equation}
    I(\Dm;\tilde{\Dm})\leq\log\rank (G)\leq S.
\end{equation}

\subsection{Information extraction from quantum memory}\label{subsec:info-extraction}
We formalize the notion of extracting information from the memory of a quantum query algorithm. The concept is purely information-theoretic as there are no computational assumptions on the extractor. We then prove that the optimal success probability of quantum memory extraction is nicely characterized by an associated potential function that we define using Compressed Oracle Duality.

\begin{definition}[Quantum memory extraction game]
    Let $\Rs$ be a fixed relation. An extractor receives the memory of $\Alg$ at time $t$, applies an arbitrary POVM, and then, without making any further queries, outputs a set of (distinct) certificates $C=\{c_1, c_2, \ldots, c_k\}\subseteq \Rs$ where $|C|=k$ (or $\bot$). The challenger measures the compressed database to obtain $\Dm = D$, and the extractor succeeds if they do not output $\bot$ and their output list satisfies $\forall c\in C: c\subseteq D$. Let $p_{\Rs,k}^*$ be the optimal success probability over all extractors.
\end{definition}

We now give a class of progress measures that one can define on the canonical representation
\[\ket{\Psi_t}
    \triangleq\sum_{\tilde{D},D}(\sqrt{G_t})_{\tilde{D},D}\ket{\tilde{D}}\ket D,
    \quad p(\tilde{D},D)\triangleq|(\sqrt{G_t})_{\tilde{D},D}|^2.\]

\begin{definition}[Joint potential]\label{def:joint-potential}
    Fix a relation $\Rs$. For each certificate $c\in \Rs$, define the projector
    \begin{equation}
        \Pi_c\triangleq \sum_{D\supseteq c}\ket D\bra D.
    \end{equation}
    The \emph{joint potential} associated with $\Rs$ is
    \begin{equation}\label{eqn:relation-progress}
        \Phi(\Rs;\Alg,t)
        \triangleq\sum_{c\in \Rs}\norm{\Pi_c\sqrt{G_t}\Pi_c}{F}^2
        =\sum_{c\in \Rs}\Tr\bigl[(\Pi_c\sqrt{G_t})^2\bigr].
    \end{equation}
    We omit $\Alg$ or $t$ when they are fixed and clear from context.
\end{definition}

Equivalently,
\begin{equation}\label{eqn:shared-certificates}
    \Phi(\Rs;\Alg,t)
    =\E_{\tilde{\Dm},\Dm}\big|\{c\in \Rs: c\subseteq \tilde{\Dm}\cap \Dm\}\big|,
\end{equation}
where the expectation is over the joint database distribution.
Note that the joint potential is generally strictly smaller than the associated standard potential used in the Compressed Oracle method; it requires certificates to exist in both the compressed oracle and the dual compressed oracle, instead of just the former. Combined with the relation to memory extraction, this is (in principle) a stronger method for proving time-space tradeoffs than the standard Compressed Oracle method.

Two important examples of joint potentials we consider are:
\begin{enumerate}
    \item \emph{Points.} Let $\Rs_{\mathrm{point}}\triangleq\{\{(x,y)\}:x,y\in[N]\}$. Then
    \begin{equation}\label{eq:point-relation}
        \Phi(\Rs_{\mathrm{point}})=\E|\tilde{\Dm}\cap\Dm|.
    \end{equation}
    Here both the input $x$ and the output $y$ must agree between the oracle and its dual.
    \item \emph{Collisions.} Let $\Rs_{\mathrm{collision}}
        =\{\{(x_1,y), (x_2,y)\}:x_1,x_2,y\in[N],\ x_1\neq x_2\}$. The potential $\Phi(\Rs_{\mathrm{collision}})$ counts the expected number of shared collisions between the compressed oracle and its dual.
\end{enumerate}

The following information extraction bound is analogous to the standard square-root measurement bound of Barnum--Knill \cite{barnumknill2002}; we give a direct proof in the specific setting of our framework.

\begin{lemma}[Potential upper bounds success probability]\label{lemma:extraction-upper}
    For every finite relation $\Rs$ and integer $1 \leq k\leq |\Rs|$, the optimal success probability $p^*_{\Rs,k}$ of outputting $k$ distinct certificates in $\Rs$ that are all contained by the database is upper bounded as
    \begin{equation}
        (p_{\Rs,k}^*)^2 \leq \frac{\Phi(\Rs)}{k}.
    \end{equation}
\end{lemma}
\begin{proof}
    Consider an optimal extractor applying POVM $\{F_C\}_{C\in\binom{\Rs}{k}}$ indexed by size-$k$ sets of certificates, which is applied to the memory of $\Alg$. We may assume that the extractor always outputs a size-$k$ set, since replacing $\bot$ by any fixed such set cannot decrease its success probability. Through Compressed Oracle Duality, we pull this back to a POVM on the image of $G_t$ by taking the conjugation $U^\dagger F_C U$ for each $F_C$. This treats the POVM as acting on the dual compressed oracle register of the canonical representation
    \[
        \ket{\Psi_t} \triangleq \sum_{\tilde{D}, D}(\sqrt{G_t})_{\tilde{D},D}\ket{\tilde{D}}\ket{D},
    \]
    since $\sqrt{G_t} = U^\dagger A_t$. We then define positive operators $E_c$ as the sum of all POVM operators containing $c$, i.e.,
    \begin{equation}
        E_c \triangleq \sum_{C\ni c} U^\dagger F_C U,
    \end{equation}
    so $0\preceq E_c \preceq \mathbb{I}$. Apply the POVM onto the dual compressed oracle $\tilde{D}$ of the canonical representation to obtain $C$. The expected number of certificates in the output set $C$ which are contained by the database $D$ is
    \begin{equation}
        \E\big|\{c\in C: c \subseteq D\}\big|=\sum_{c\in \Rs}\Tr(E_c\sqrt{G_t}\Pi_c\sqrt{G_t})=\sum_{c\in \Rs}\Tr\bigl[(G_t^{1/4}E_cG_t^{1/4})(G_t^{1/4}\Pi_cG_t^{1/4})\bigr].
    \end{equation}
    Cauchy--Schwarz for the direct sum of Hilbert--Schmidt inner products gives
    \begin{equation}\label{eq:squared-size-bound}
        \left(\E\big|\{c\in C: c \subseteq D\}\big|\right)^2\leq
        \left(\sum_{c\in \Rs}
          \Tr\left[\big(G_t^{1/4}E_cG_t^{1/4}\big)^2\right]\right)
        \Phi(\Rs).
    \end{equation}
    Since $G_t^{1/4}E_cG_t^{1/4} \preceq \sqrt{G_t}$,
    \begin{equation}
        \sum_{c\in \Rs}
          \Tr\left[\big(G_t^{1/4}E_c G_t^{1/4}\big)^2\right] \leq \sum_{c\in \Rs} \Tr\left[E_c G_t\right] = \E[|C|] = k.
    \end{equation}
    Plugging this into \eqref{eq:squared-size-bound} and using the fact that success requires all $k$ certificates contained in $D$, we get
    \begin{equation}
        k^2 (p_{\Rs,k}^*)^2 \leq  \left(\E\big|\{c\in C: c \subseteq D\}\big|\right)^2 \leq k\Phi(\Rs).
    \end{equation}
    Rearranging terms completes the proof.
\end{proof}

For the extraction probability lower bound, we require the following definition:
\begin{equation}
    \lambda(\Rs;\Alg,t)\triangleq
    \max_{\substack{c\in \Rs\\\Pr[c\subseteq \tilde{\Dm}\cap\Dm]>0}}\,
    \E\left[|\{c'\in \Rs:c'\subseteq\tilde{\Dm}\}|
        \,\middle|\,c\subseteq\tilde{\Dm}\cap\Dm\right],
\end{equation}
where the probability and expectation are taken over the joint database distribution $p(\tilde{D}, D)$. In words, it is the maximum expected number of certificates in the dual compressed oracle, conditioned on a certificate being shared by both oracles. We set $\lambda(\Rs;\Alg,t)=1$ when the max is over the empty set, and omit $\Alg$ or $t$ when they are fixed and clear from context.

\begin{lemma}[Canonical extraction lower bounds success probability]
    \label{lemma:extraction-lower}
    The optimal success probability $p^*_{\Rs,1}$ of outputting a single
    certificate in $\Rs$ that is contained in the database $D$ is lower
    bounded as
    \begin{equation}
        p_{\Rs,1}^* \geq \frac{\Phi(\Rs)}{\lambda(\Rs)}.
    \end{equation}
\end{lemma}
\begin{proof}
    The claim is immediate when $\Phi(\Rs)=0$, so suppose $\Phi(\Rs)>0$. We give an explicit POVM achieving at least the claimed lower bound probability. This POVM is constructed by applying the partial-isometry $U^\dagger$ constructing the dual compressed oracle, then measuring to obtain $\tilde{\Dm}$. For each certificate $c'\in \Rs$ contained in $\tilde{\Dm}$, output $C=\{c'\}$ with probability $1/|\{c\in \Rs:c\subseteq\tilde{\Dm}\}|$, and output $\bot$ if there are no such certificates. This is a valid measurement because the output probabilities sum to one. Its success probability is
    \begin{equation}
        \E\frac{|\{c\in \Rs:c\subseteq\tilde{\Dm}\cap\Dm\}|}
                 {|\{c\in \Rs:c\subseteq\tilde{\Dm}\}|}\geq
            \frac{\Phi(\Rs)^2}{
                \E\left[|\{c\in \Rs:c\subseteq\tilde{\Dm}\}|\,|\{c\in \Rs:c\subseteq\tilde{\Dm}\cap\Dm\}|
                \right]}\geq
            \frac{\Phi(\Rs)}{\lambda(\Rs)}.
    \end{equation}
    The fraction inside the expectation is defined to be zero when its denominator is zero. The first inequality follows from Cauchy--Schwarz and the definition of the joint potential. The second follows by expanding $|\{c\in \Rs:c\subseteq\tilde{\Dm}\cap\Dm\}|$ as a sum of indicators and applying the definition of $\lambda$ to each term.
\end{proof}

\cref{lemma:extraction-upper,lemma:extraction-lower} show that the
success of the optimal memory extraction of an arbitrary relation
from a compressed oracle database is characterized by its associated
joint potential:
\begin{equation}
    \frac{\Phi(\Rs)}{\lambda(\Rs)}
    \leq p_{\Rs,1}^*
    \leq \sqrt{\Phi(\Rs)}.
\end{equation}

Now applying the upper bound to the original random oracle $\Hm$ for single-certificate output, we get:
\begin{corollary}\label{corollary:extraction-progress}
    Let $\Rs$ be a relation with certificates of size $\ell$. Consider a quantum adversary querying a uniform random oracle $\Hm$. Let $\ket{\psi_t^{\Hm}}$ be the adversary's memory after $t$ queries. Suppose the adversary applies any arbitrary measurement on $\ket{\psi_t^{\Hm}}$ and outputs $c\in \Rs$ satisfying $c\subseteq \Hm$ with probability $p$. Then
    \begin{equation}
        \sqrt{p}\leq\Phi(\Rs;\Alg,t)^{1/4}+\sqrt{\ell/N}.
    \end{equation}
\end{corollary}
\begin{proof}
    Let $p_\Rs$ be the probability the certificate output $c$ is contained in the database of the compressed oracle simulation of the algorithm. By \cref{lemma:compressed_oracle,lemma:extraction-upper},
    \begin{equation}
        \sqrt p\leq\sqrt{p_\Rs}+\sqrt{\ell/N}
        \leq\Phi(\Rs;\Alg,t)^{1/4}+\sqrt{\ell/N}.
    \end{equation}
\end{proof}

In particular, if an adversary outputs a single valid collision $(x_1, x_2, y)$ with probability $p$, then
\begin{equation}\label{eqn:collision-endpoint}
    \Phi(\Rs_{\mathrm{collision}};\Alg,t)
    \geq \left(\max\left\{0,\sqrt p-\sqrt{2/N}\right\}\right)^4.
\end{equation}

\subsection{A candidate approach towards collision time-space tradeoffs}\label{subsec:candidate-approach}
We now give a candidate procedure (that ultimately does not give an unconditional result) for how one might prove quantum time-space tradeoffs for collision finding \emph{without} assuming label symmetry. Intuitively, progress towards finding a collision is determined by the number of points you can store in your memory: for example, BHT \cite{Brassard_1998} gives a quantum speedup by storing $\Omega(N^{1/3})$ points and utilizing coherent amplitude amplification on the entire set. 

Motivated by this, we use Compressed Oracle Duality to prove that an adversary utilizing $S$ qubits of space can generally only output approximately $S/\log N$ points from a random oracle.

\begin{lemma}[Shared-point bound]\label{lemma:point-extraction}
    For $1\leq t<N$,
    \begin{equation}\label{eqn:point-space}
        \Phi(\Rs_{\mathrm{point}};\Alg,t)
        \leq\left(\sqrt{\frac{S+\log e}{\log N-\log t}}+\sqrt{\frac{t}{N}}\right)^2.
    \end{equation}
    In particular, if $S\geq \log N$ and $t\leq \sqrt{N}$, then $\Phi(\Rs_{\mathrm{point}};\Alg,t)=\Os(S/\log N)$.
\end{lemma}
\begin{proof}
    Measure the dual compressed oracle to obtain a database $\tilde{\Dm}$, and decompress and measure the original compressed oracle to obtain the full oracle $\Hm$. These operations commute. Equivalently, $\Hm$ is uniform and $\tilde{\Dm}$ is obtained by a fixed POVM on the adversary's memory for that oracle. Moreover, $|\supp(\tilde{\Dm})|\leq t$.

    For each fixed dual compressed oracle $\tilde{D}$,
    \begin{equation}
        \E_{\Hm}(N/t)^{|\Hm\cap\tilde{D}|}
        = \left(1+\frac{N/t-1}{N}\right)^{|\tilde{D}|}\leq(1+1/t)^t\leq e.
    \end{equation}
    We apply the log-sum inequality to the conditional law of $\Hm$ given each dual compressed oracle outcome $\tilde{D}$ and the uniform prior. Averaging over $\tilde{\Dm}$ gives
    \begin{equation}
        (\log N-\log t)\E\big[|\Hm\cap\tilde{\Dm}|\big]-\log e \leq I(\Hm;\tilde{\Dm}).
    \end{equation}
    By \cref{lem:holevo-bound}, $I(\Hm;\tilde{\Dm})\leq S$, since the memory dimension is at most $2^S$. Therefore,
    \begin{equation}\label{eqn:point-readout}
        \E\big[|\Hm\cap\tilde{\Dm}|\big]\leq\frac{S+\log e}{\log N-\log t}.
    \end{equation}

    For points, consider the extractor that outputs each entry of $\tilde{\Dm}$ with probability $1/t$, and otherwise fails. The success probabilities against the compressed oracle database and the true oracle are $\Phi(\Rs_{\mathrm{point}})/t$ and $\E\big[|\Hm\cap\tilde{\Dm}|\big]/t$, respectively. Applying \cref{lemma:compressed_oracle} with $\ell=1$ gives
    \begin{equation}\label{eqn:decompression-loss}
        \left|\sqrt{\E\big[|\Hm\cap\tilde{\Dm}|\big]}-\sqrt{\Phi(\Rs_{\mathrm{point}})}\right|
        \leq\sqrt{t/N}.
    \end{equation}
    Combining \eqref{eqn:point-readout} and \eqref{eqn:decompression-loss} proves \eqref{eqn:point-space}. Finally, $\log N -\log t\geq \log N/2$ when $t\leq \sqrt{N}$, which gives the stated $\Os(S/\log N)$ bound.
\end{proof}

Both point and collision potentials start at zero, and oracle-independent memory unitaries leave them unchanged. \cref{lemma:point-extraction} bounds the number of extractable points a bounded quantum memory can store at any time, while \eqref{eqn:collision-endpoint} gives a joint potential that, if bounded, would imply collision time-space tradeoffs. Unfortunately, a full unconditional result requires an additional bound on progress of the joint potential, which \cref{lemma:point-extraction} by itself does not imply\footnote{Intuitively, one might expect the collision progress to be bounded in relation to the number of points that are extractable from memory. However, in certain examples this turns out to not be exactly true.}. In the following section, we show how label symmetry is able to bypass this barrier using a similar idea but without relying on joint potentials.

%% file: quantum-label.tex
\section{A simplified proof for label-symmetric quantum algorithms}\label{sec:label-symmetry}
In this section, we apply Compressed Oracle Duality to obtain a simplified proof of the quantum time-space tradeoff assuming label symmetry from~\cite{magniez2026label}. Let $\mathfrak{S}_N$ denote the symmetric group on $N$ elements. We consider the label-permutation action on the set of oracles $H:[N]\rightarrow [N]$ defined by
\begin{equation}\label{eq:symmetric-group-action}
    H \mapsto \sigma\circ H, \qquad(\sigma\circ H)(x)\triangleq \sigma(H(x)).
\end{equation}
The label-permutation action extends to compressed oracle databases by fixing $\sigma(\bot) = \bot$ for all $\sigma\in\mathfrak{S}_N$, and it commutes with the $\StdDecomp$ operator of the compressed oracle.

\begin{definition}[Label symmetry assumption {\cite[Fact~4.2 and Definition~4.4]{magniez2026label}}]\label{def:label-symmetry}
    Write $\ket{\psi_t^H}$ for the quantum adversary state with oracle $H$ after $t$ queries. The algorithm is \emph{label-symmetric} if, for every $t$, every pair of oracles $H_1,H_2$, and every permutation $\sigma \in \mathfrak{S}_N$,
    \begin{equation}
        \braket{\psi_t^{H_1}|\psi_t^{H_2}}
        =\braket{\psi_t^{\sigma\circ H_1}|\psi_t^{\sigma\circ H_2}}.
    \end{equation}
    This is equivalently stated as the Gram matrix $G$ satisfies $G_{D_1, D_2} = G_{\sigma\circ D_1, \sigma\circ D_2}$ for all databases $D_1, D_2$ in the compressed-oracle simulation of the algorithm.
\end{definition}

The main result of Magniez--Zur \cite{magniez2026label} proves a quantum time-space tradeoff of $T^2S = \Omega(N\log N)$ assuming label symmetry:
\begin{theorem}[Quantum time-space tradeoff {\cite[Theorem~4.19]{magniez2026label}}]\label{thm:label-quantum}
    Any label-symmetric quantum algorithm that outputs a triple $(x_1,x_2,y)$ satisfying $x_1<x_2$ and $\Hm(x_1)=\Hm(x_2)=y$ with probability at least $2/3$, using $T$ queries and $S$ qubits, satisfies
    \begin{equation}
        T^2S=\Omega(N\log N).
    \end{equation}
\end{theorem}

Their proof utilizes the representation theory of the symmetric group action on the space of valid compressed oracles. We give a brief overview of the Magniez--Zur proof in \cref{subsec:mz-proof}. We then give an elementary alternative proof in \cref{subsec:label-quantum} which does not require representation theory and instead bounds the mutual information between the two marginals of the joint database distribution.

\subsection{Bounding progress via collision-free capacity}\label{subsec:mz-proof}
We begin by bounding a collision-finding algorithm’s progress in terms of a quantity called ``collision-free capacity.'' This standard bound is implicit in the proof of the space-agnostic collision lower bound using compressed oracles (e.g.,~\cite{zhandry2019record,hamoudi2023quantum}) and is the starting point for the proof in~\cite{magniez2026label} (as well as our simplified proof). Note that our notation differs slightly from~\cite{magniez2026label}.

We define the ``collision-free capacity'' as follows:
\begin{definition}[Collision-free capacity]
    For a Gram matrix $G$, we define its collision-free capacity as the expected database size restricted to collision-free outcomes:
    \begin{equation}\label{eq:label-collision-capacity}
        \Cs(G)\triangleq\sum_{D:\CF(D)}|D|(G)_{D,D}.
    \end{equation}
\end{definition}

\paragraph{Standard compressed oracle setup.} Let $\ket{\psi_t}$ be the joint adversary-oracle state after $t$ queries. Define the projectors $\Pi_{=0}$ and $\Pi_{\geq 1}$ which project onto the space of compressed databases without and with a collision, respectively, and $\Lambda_s$ which projects onto databases $D$ where $|D|=s$. Define the progress measure
\[\Delta_t\triangleq \|\Pi_{\geq 1}\ket{\psi_t}\!\|,\]
which has the per-query bound $\Delta_{t+1} - \Delta_t \leq \|\Pi_{\geq 1}\CStO\, \Pi_{=0} \ket{\psi_t}\!\|$ when $\ket{\psi_t}$ is the normalized joint adversary-oracle state. Magniez--Zur (as well as the original collision lower bound from~\cite{zhandry2019record}) gives
\begin{equation}
    \|\Pi_{\geq 1}\CStO\, \Pi_{=0} \ket{\psi_t}\!\| \leq \sqrt{\frac{16\Cs(G_t)}{N}}.
\end{equation}
Applying \cref{lemma:compressed_oracle}, Magniez--Zur obtain the bound on success probability $p$ of outputting a collision using $T$ queries as
\begin{equation}
    p \leq \left(\Delta_T + \sqrt{\frac{2}{N}}\right)^2.
\end{equation}
Putting these together shows that proving lower bounds reduces to proving an upper bound on $\Cs(G_t)$. The query lower bound of~\cite{zhandry2019record} assumes no bound on space and simply uses $\Cs(G_t)\leq t$ to immediately get $p= \Os(\frac{T^3}{N})$ for $T\geq1$, or $T = \Omega(N^{1/3})$ to succeed with constant probability. Magniez--Zur, by additionally assuming a space bound $S$ and label symmetry, are able to prove stronger lower bounds with an improved bound on $\Cs(G_t)$.

\paragraph{Magniez--Zur proof of the collision-free capacity bound.} We briefly summarize how Magniez--Zur prove their capacity bound, assuming label symmetry. The reader can also safely skip to \cref{subsec:label-quantum} to see our simplified proof. 

The main technical contribution of Magniez--Zur is to show that the squared amplitude on collision-free databases of size $s$ in the compressed oracle, for $0\leq s\leq t\leq\sqrt N$, satisfies:
\begin{equation}\label{eq:mz-capacity-lemma}
    \|\Pi_{= 0}\Lambda_s\ket{\psi_t}\!\|^2 \leq \begin{cases} 
      \|\Lambda_s\ket{\psi_t}\!\|^2 & s\leq \frac{2S}{\log N},\\
     \frac{2s-2}{N} \|\Lambda_s\ket{\psi_t}\!\|^2 & s> \frac{2S}{\log N}.
   \end{cases}
\end{equation}
Taking the summation over all $s$ weighted by squared amplitude, and using $s(2s-2)/N\leq2\leq2S/\log N$ in the second case, gives
\begin{equation}
    \Cs(G_t) = \sum_{s=0}^t s\|\Pi_{= 0}\Lambda_s\ket{\psi_t}\!\|^2 \leq \min\{t, 2S/\log N\},
\end{equation}
thereby achieving the optimal tradeoff of $T^2S=\Omega(N\log N)$ assuming label symmetry.

To prove the collision-free capacity bound in \eqref{eq:mz-capacity-lemma}, fix the $s$ occupied database positions and omit the remaining cells. Define $W\triangleq\ker(\bra{\hat{0}})\subseteq\mathbb C^N$ as the nonempty part of one valid compressed cell. Since $\StdDecomp$ swaps $\ket{\hat{0}}$ and $\ket{\bot}$, validity on these positions corresponds to $W^{\otimes s}$. Additionally, Magniez--Zur define the collision-free subspace $C_s^\perp$ as the span of tuples with pairwise distinct labels.

The proof studies $W^{\otimes s}\cap C_s^\perp$ under two commuting actions: $\mathfrak S_N$ relabels entries, while $\mathfrak S_s$ permutes positions. Projections constructed from position permutations split $C_s^\perp$ into $\mathfrak S_N$-invariant subspaces, which are then decomposed into irreducible representations called Specht modules $S^\lambda$. These are indexed by partitions $\lambda\vdash N$, whose parts give the row lengths of the corresponding Young diagrams. Representation-theoretic analysis of a symmetric-group object called the arrangement-graph shows that, for $N\geq 2s$, the intersection consists precisely of copies of $S^{(N-s,\mu)}$, where $\mu\vdash s$ \cite[Theorem~4.6 and Section~6]{magniez2026label}. The memory bound on the spans of relabeled states rules out the intersection component when these modules have sufficiently large dimension \cite[Theorem~4.7 and~4.15]{magniez2026label}, and the arrangement-graph's spectral gap bounds control the remaining collision-free amplitude \cite[Lemma~4.11 and Theorem~4.15]{magniez2026label}.

\subsection{A simplified proof of the collision-free capacity bound}
\label{subsec:label-quantum}
We now give our alternative proof of the collision-free capacity bound which does not rely on the representation theory tools used in \cite{magniez2026label}. 

Call a trace-one Gram matrix $G$ on compressed databases \emph{label-symmetric} if $G_{\sigma\circ D_1,\sigma\circ D_2}=G_{D_1,D_2}$ for all databases $D_1,D_2$ and all $\sigma\in\mathfrak{S}_N$. The Gram matrix $G_t$ and its square-root $\sqrt{G_t}$ of a label-symmetric algorithm is label-symmetric by standard compression identities~\cite[Sections~3.1 and~4.1]{magniez2026label}. 

Let $p(\tilde{D},D)\triangleq|(\sqrt{G})_{\tilde{D},D}|^2$ be the joint database distribution. We split the labels of a collision-free oracle $D$ into those present in $\tilde{D}$ and those absent. Applying the Holevo bound (\cref{lem:holevo-bound}) and label symmetry to the joint database distribution bounds the amplitude on matched labels; symmetry and validity bound the unmatched labels. This recovers \cite[Corollary~4.16]{magniez2026label} up to constant factors, and the remainder of the proof essentially follows Magniez--Zur.
\begin{lemma}[Quantum collision-free capacity]\label{lem:quantum-label-capacity}
    Let $0\leq t\leq\sqrt N$, and let $G$ be a trace-one, valid, label-symmetric Gram matrix that is non-zero only on entries $G_{D_1, D_2}$ with $|D_1|, |D_2|\leq t$. Then
    \begin{equation}\label{eq:quantum-label-capacity}
        \Cs(G)\leq \min\left\{t,\frac{4\log\rank(G)}{\log N}+\frac{8t^2}{N}\right\}.
    \end{equation}
    In particular, $\Cs(G)\leq\min\{t,12S/\log N\}$ whenever $\rank(G)\leq2^S$.
\end{lemma}
\begin{proof}
    The case $t=0$ is immediate. We omit the fixed $G$ argument from $\Cs$. Let $p_D\triangleq G_{D,D}$ be the standard database distribution and let $p(\tilde{D},D)\triangleq|(\sqrt{G})_{\tilde{D},D}|^2$ denote the joint database distribution. Define the matched and unmatched label counts
    \begin{equation}
        \begin{aligned}
            \Cs^{+}
            &\triangleq \E\big[\id_{\CF(\Dm)}\cdot |\range(\Dm)\cap\range(\tilde{\Dm})|\big],\\
            \Cs^{-}
            &\triangleq \E\big[\id_{\CF(\Dm)}\cdot |\range(\Dm)\setminus\range(\tilde{\Dm})|\big],
        \end{aligned}
    \end{equation}
    where the expectations are over the joint database distribution.
    Thus $\Cs^{+}$ is similar to the expectation defining $\Phi(\Rs_{\mathrm{point}})$ in \cref{subsec:info-extraction}, but matched labels may occur at different inputs, and the expectation is restricted to collision-free databases $D$. The dual compressed oracle database $\tilde{D}$ need not be collision-free. Since the values of a collision-free $D$ are distinct,
    \begin{equation}\label{eq:quantum-label-capacity-decomposition}
        \Cs= \E[\id_{\CF(\Dm)}\cdot |\range(\Dm)|] = \Cs^{+}+\Cs^{-},
    \end{equation}
    and we bound both contributions separately.

    \paragraph{Matched labels.}
    Consider a database distribution $\Dm$, and condition on $\CF(\Dm)$ and a fixed support $X$ of size $s$ for the entire matched-label calculation. Label symmetry makes $\Dm$ uniform over the $N!/(N-s)!$ injective labelings of $X$, so $\mathrm{H}(\Dm)=\log(N!/(N-s)!)$ below.
    
    Let $L(D,\tilde{D}) \triangleq \range(D)\cap\range(\tilde{D})$ denote the set of matched labels between $D$ and $\tilde{D}$. Since $|\range(\tilde{D})|\leq t$, any specified size-$k$ subset of $X$ all map to labels within $\range(\tilde{D})$ with
    probability at most $(t/N)^k$. Summing over these size-$k$ subsets of $X$ gives
    \begin{equation}
        \E_{\Dm}\binom{|L(\Dm,\tilde{D})|}{k}
        \leq \binom{s}{k}\left(\frac{t}{N}\right)^k.
    \end{equation}
    The binomial theorem gives
    \begin{equation}
        \E_{\Dm}\left[\left(\frac{N}{t}\right)^{|L(\Dm,\tilde D)|}\right]=\sum_{k=0}^s\left(\frac{N}{t}-1\right)^k \E_{\Dm}\binom{|L(\Dm,\tilde D)|}{k}\leq \left(2-\frac{t}{N}\right)^s\leq 2^s,
    \end{equation}
    where $\tilde{D}$ is fixed and the expectation is taken over the uniform marginal of $\Dm$. Applying the log-sum inequality to the conditional law of $\Dm$
    given $\tilde{\Dm}=\tilde{D}$ gives
    \begin{equation}
        \mathrm{H}(\Dm\mid\tilde{\Dm}=\tilde{D}) +\log(N/t)\,\E\left[|L(\Dm,\tilde D)|\middle| \tilde{\Dm}=\tilde {D}\right]\leq\log\sum_{\substack{D:\supp(D)=X\\\CF(D)}}(N/t)^{|L(D,\tilde D)|}\leq \mathrm{H}(\Dm)+s.
    \end{equation}
    Averaging over the dual compressed oracle value $\tilde{D}$ therefore yields
    \begin{equation}\label{eq:label-counting}
        \mathrm{H}(\Dm\mid\tilde{\Dm})
        \leq \mathrm{H}(\Dm)+s-\log(N/t)\,
        \E\big[|\range(\Dm)\cap\range(\tilde{\Dm})|\big].
    \end{equation}
    
    Returning to the unconditioned joint distribution, since $\supp(\Dm)$ and $\id_{\CF(\Dm)}$ are functions of $\Dm$, the chain rule gives
    \begin{equation}
        \begin{aligned}
        I(\Dm;\tilde{\Dm})&= I\big(\supp(\Dm),\id_{\CF(\Dm)};\tilde{\Dm}\big)+I\big(\Dm;\tilde{\Dm}\mid\supp(\Dm),\id_{\CF(\Dm)}\big)\\
        &\geq \sum_{X\subseteq[N]}\Pr[\supp(\Dm)=X,\ \CF(\Dm)]\cdot I\big(\Dm;\tilde{\Dm}\mid\supp(\Dm)=X,\ \CF(\Dm)\big),
        \end{aligned}
    \end{equation}
    where we dropped the first term and the conditional terms with a collision, all of which are nonnegative. By \eqref{eq:label-counting}, each remaining term is at least
    \begin{equation}
        \log(N/t)\,\E\left[|\range(\Dm)\cap\range(\tilde{\Dm})|\middle|\supp(\Dm)=X,\ \CF(\Dm)\right]-|X|.
    \end{equation}
    The summation then gives
    \begin{equation}\label{eq:label-rank-readout}
        \log(N/t)\,\Cs^+-\Cs \leq I(\Dm;\tilde{\Dm}) \leq \log\rank(G),
    \end{equation}
    where the last inequality follows from \cref{lem:holevo-bound}.
    
    \paragraph{Unmatched labels.} We bound $\Cs^{-}$ using label symmetry and validity. Fix a collision-free database $D_0$ with $|D_0|\leq t-1$, an input $x\notin\supp(D_0)$, and a database $\tilde{D}$ with $|\tilde{D}|\leq t$. Define the set $B\triangleq\range(D_0)\cup\range(\tilde{D})$. Every permutation fixing $B$ pointwise fixes both $D_0$ and $\tilde{D}$, so label symmetry makes the value $(\sqrt{G})_{\tilde{D},D_0\cup\{(x,y)\}}$ constant for $y\notin B$, which are the labels absent from both $D_0$ and the dual $\tilde{D}$. Denote this constant $\alpha$. Validity also gives 
    \begin{equation}\label{eq:label-validity}
        \sum_{y\in[N]}(\sqrt{G})_{\tilde{D},D_0\cup\{(x,y)\}}=0.
    \end{equation}
    Applying Cauchy--Schwarz,
    \begin{equation}
        (N-|B|)^2|\alpha|^2 =\left|\sum_{y\in B}(\sqrt{G})_{\tilde{D},D_0\cup\{(x,y)\}}\right|^2 \leq |B|\sum_{y\in B}\left|(\sqrt{G})_{\tilde{D},D_0\cup\{(x,y)\}}\right|^2.
    \end{equation}
    Plugging in $\alpha$ and simplifying gives
    \begin{equation}\label{eq:unmatched-bound}
        \sum_{y\notin B}\left|(\sqrt{G})_{\tilde{D},D_0\cup\{(x,y)\}}\right|^2
        \leq\frac{|B|}{N}\sum_{y\in[N]}\left|(\sqrt{G})_{\tilde{D},D_0\cup\{(x,y)\}}\right|^2.
    \end{equation}
    
    Now sum over all such $x,D_0,\tilde{D}$. Each unmatched label of a collision-free $D$ is counted once on the left, by removing its unique input--output pair to obtain $D_0$. The right side may include colliding completions $D=D_0\cup\{(x,y)\}$, but each completed pair $(\tilde{D},D)$ is counted at most $|D|\leq t$ times. Since $\sum_{\tilde{D},D}|(\sqrt{G})_{\tilde{D},D}|^2=1$, we get
    \begin{equation}\label{eq:label-unmatched}
        \Cs^{-}\leq\frac{t(2t-1)}{N}\leq\frac{2t^2}{N}.
    \end{equation}
    Using $\Cs=\Cs^++\Cs^-$, $\log(N/t)\geq\log N/2$, and $\log N\geq6$ in \eqref{eq:label-rank-readout} gives $\Cs\leq4\log\rank(G)/\log N+2\Cs^-$. Together with \eqref{eq:label-unmatched} and $\Cs\leq t$, this proves \eqref{eq:quantum-label-capacity}. The final claim follows from $\log\rank(G)\leq S$, $t^2/N\leq1$, and $S\geq\log N$.
\end{proof}

We now complete the proof of the time-space tradeoff for label-symmetric quantum algorithms.
\begin{proof}[Proof of \cref{thm:label-quantum}]
    If $T>\sqrt N$, the result follows from $S\geq\log N$. Otherwise, the progress argument in the proof of \cite[Lemma~4.17]{magniez2026label} and \cref{lemma:compressed_oracle} give
    \begin{equation}\label{eq:label-total-progress}
        \sqrt{2/3} \leq \frac{4}{\sqrt N} \sum_{t<T}\sqrt{\Cs(G_t)}+\sqrt{2/N} \leq 4T\sqrt{\frac{12S}{N\log N}}+\sqrt{2/N},
    \end{equation}
    where the last inequality applies \cref{lem:quantum-label-capacity} to $G_t$, which satisfies its hypotheses with $\rank(G_t)\leq2^S$. Since $N\geq64$, this implies $T^2S=\Omega(N\log N)$.
\end{proof}

\subsection{Intermediate measurements in label-symmetric algorithms}\label{subsec:intermediate-measurements}
Magniez and Zur extend their tradeoff to algorithms with intermediate measurements under a conditional form of label symmetry~\cite[Section~5.3]{magniez2026label}. Each step of the algorithm is a Kraus map on the memory, and, following~\cite{chung2020function}, the analysis conditions on the history $h$ of the measurement outcomes. The adversary state on oracle $H$ after $t$ queries is therefore a direct sum of unnormalized pure states $\ket{\psi_{t,h}^H}$, whose squared norm is the probability of branch $h$. At every query, part of the adversary memory is a classical record register $\mathsf{R}$ with basis states indexed by set $\Omega$. $\mathsf{R}$ contains the value $z_h\in \Omega$ on branch $h$ which counts against the space bound $S$. $\mathfrak{S}_N$ also acts on $\mathsf{R}$ by relabeling whatever labels $z_h$ stores. The label symmetry condition of \cite{magniez2026label} is then defined as follows:
\begin{definition}[Label symmetry with intermediate measurements]\label{def:label-symmetry-measure}
    For an element $z\in \Omega$, let $\mathfrak{S}_{z}\subseteq\mathfrak{S}_{N}$ denote the stabilizer of $z$: 
    \begin{equation}
        \mathfrak{S}_{z}\triangleq\{\sigma\in\mathfrak{S}_N:\sigma(z)=z\}.
    \end{equation}
    An algorithm with intermediate measurements is label-symmetric if, for every $t$, every pair of oracles $H_1,H_2$, every branch $h$, and every permutation $\sigma \in \mathfrak{S}_{z_h}$,
    \begin{equation}
        \braket{\psi_{t,h}^{H_1}|\psi_{t,h}^{H_2}}
        =\braket{\psi_{t,h}^{\sigma\circ H_1}|\psi_{t,h}^{\sigma\circ H_2}}.
    \end{equation}
\end{definition}

Let $\ket{\psi_{t,h}}$ denote the unnormalized joint adversary-oracle state on branch $h$ under a uniformly random oracle. The new progress measure is defined in \cite[Section~5.3]{magniez2026label} as
    \begin{equation}
        \Delta_t\triangleq \left(\sum_{h} \|\Pi_{\geq 1}\ket{\psi_{t,h}}\!\|^2\right)^{1/2}
    \end{equation}

With this definition, the proof of \cref{thm:label-quantum} extends fairly easily:
\begin{corollary}\label{cor:intermediate-measurements}
    \Cref{thm:label-quantum} also holds for algorithms with intermediate measurements that satisfy \cref{def:label-symmetry-measure}.
\end{corollary}
\begin{proof}
    Let $G_{t,h}$ be the Gram matrix of $\ket{\psi_{t,h}}$, so that $\sum_h\Tr(G_{t,h})=1$. The Kraus operators act only on the memory, so they commute with the projectors $\Pi_{=0}$ and $\Pi_{\geq1}$ on the compressed database, and the progress and success bounds of \cref{subsec:mz-proof} hold for every measurement branch~\cite[Section~5.3]{magniez2026label}. This gives
    \begin{equation}
        \Delta_{t+1}\leq\Delta_t+\sqrt{\frac{16}{N}\sum_h\Cs(G_{t,h})}, 
    \end{equation}
    and the success probability is still at most $(\Delta_T+\sqrt{2/N})^2$. As in the proof of \cref{thm:label-quantum}, it therefore suffices to show that $\Cs(G_{t,h})\leq(12S/\log N)\Tr(G_{t,h})$ for every $t\leq\sqrt N$ and every branch $h$.
    
    Fix a branch $h$ that occurs with nonzero probability, and write the normalized branch state $\ket{\psi_{t,h}}/\|\ket{\psi_{t,h}}\|$ as $\ket{z_h}_{\mathsf{R}}\ket{\psi_h}_{\mathsf{A\setminus R, D}}$, with Gram matrix $G_h\triangleq G_{t,h}/\Tr(G_{t,h})$. The record register $\mathsf{R}$ has dimension $|\Omega|$ and is fixed in the branch, so $\rank(G_h)\leq 2^S/|\Omega|$. By the label symmetry condition, $G_h$ is invariant under relabeling by $\mathfrak{S}_{z_h}$, but not necessarily by all of $\mathfrak{S}_N$, so \cref{lem:quantum-label-capacity} does not apply to it directly.
    
    To deal with this, we ``label-symmetrize'' $G_h$ by twirling it over the symmetric group! Let $V_\sigma\ket{D}\triangleq\ket{\sigma\circ D}$ denote the relabeling action of $\sigma$ on the compressed oracle, and let $\mathfrak{S}_{h}\subseteq\mathfrak{S}_N$ denote the set which contains one representative from each left coset in $\mathfrak{S}_N/\mathfrak{S}_{z_h}$. Then the records $\sigma(z_h)$ for $\sigma\in\mathfrak{S}_{h}$ are distinct, and $|\mathfrak{S}_{h}|\leq|\Omega|$. Consider the state
    \begin{equation}
        \ket{\phi}\triangleq\frac{1}{\sqrt{|\mathfrak{S}_{h}|}}\sum_{\sigma\in\mathfrak{S}_{h}}\ket{\sigma(z_h)}_{\mathsf{R}}\otimes(\mathbb{I}\otimes V_\sigma)\ket{\psi_h}_{\mathsf{A\setminus R,D }}.
    \end{equation}
    The states $\ket{z}$ for each record $z\in\Omega$ are orthonormal, so the Gram matrix of $\ket{\phi}$ is 
    \[|\mathfrak{S}_{h}|^{-1}\sum_{\sigma\in\mathfrak{S}_{h}}V_\sigma G_h V_\sigma^\dagger.\]
    Since $G_h$ is $\mathfrak{S}_{z_h}$-invariant, each term depends only on the coset of $\sigma$, so this equals the twirl
    \[G_\phi=\frac{1}{|\mathfrak{S}_N|}\sum_{\sigma\in\mathfrak{S}_N}V_\sigma G_h V_\sigma^\dagger.\]
    
    We verify the hypotheses of \cref{lem:quantum-label-capacity} for $G_{\phi}$. It has trace-one, and it is label symmetric because it averages over the whole group. It is valid because relabeling fixes $\ket\bot$ and $\ket{\hat{0}}$ and so maps the valid subspace to itself, and it is supported on databases with at most $t$ entries because relabeling preserves database sizes. Its rank is at most $|\mathfrak{S}_{h}|\cdot\rank(G_h)\leq|\Omega|\cdot2^S/|\Omega|=2^S$. Relabeling also preserves collision-freeness, so $\Cs(G_{\phi})=\Cs(G_{h})$, and \cref{lem:quantum-label-capacity} gives $\Cs(G_{h})\leq 12S/\log N$. Multiplying by $\Tr(G_{t,h})$ proves the claim.
\end{proof}

The label-symmetrization argument is the counterpart of the orbit-span bound in~\cite[Section~5.3]{magniez2026label}: it realizes the $\mathfrak{S}_N$-orbit of a branch as a single, fully label-symmetric pure state.

%% file: classical-label.tex
\section{Time-space tradeoff for label-symmetric classical algorithms}\label{sec:label-classical}
In this section we prove the time-space tradeoff for deterministic\footnote{This also includes randomized algorithms which are label-symmetric for each random seed.}, potentially non-reversible, label-symmetric classical algorithms.

\begin{definition}
    A deterministic classical algorithm, with memory state $\mu_t(H)$ on oracle $H$, is label-symmetric if
    \begin{equation}\label{eq:ls-classical-symmetry}
        \mu_t(H_1)=\mu_t(H_2)
        \Longleftrightarrow
        \mu_t(\sigma\circ H_1)=\mu_t(\sigma\circ H_2)
    \end{equation}
    for every $t,H_1,H_2,\sigma$.
\end{definition}
\begin{remark}
    The classical condition is the quantum condition with distinct memory states represented by distinct computational basis vectors. 
\end{remark}

Our time-space tradeoff is $TS=\Omega(N\log N)$, where we note again that Pollard's rho algorithm is a deterministic algorithm achieving $TS = \Os(\sqrt{N}\log N)$; this implies that label symmetry may be an overly-restrictive assumption.
\begin{theorem}[Classical time-space tradeoff]\label{thm:classical-label-tradeoff}
    Any label-symmetric deterministic classical algorithm that outputs a triple $(x_1,x_2,y)$ satisfying $x_1<x_2$ and $\Hm(x_1)=\Hm(x_2)=y$ with probability at least $2/3$, using $T$ queries and $S$ bits, satisfies
    \begin{equation}
        TS=\Omega(N\log N).
    \end{equation}
\end{theorem}

Our proof is again inspired by Compressed Oracle Duality; it is modeled after the quantum proof by identifying the set of ``matched labels'' for a classical memory state. We again use duality and the Holevo bound to bound the information in these labels; however, we note the same bound also follows from a direct memory-state counting argument. The main difference between the classical and quantum proofs is that we bound classical probability progress by directly comparing with the computation path of an injective function (with no collisions). 

We model computation using an $R$-way branching program, introduced by Borodin and Cook~\cite{BorodinCook1982}. Branching programs represent any arbitrary computation using bounded memory and query complexity. This model was used, for example, by Dinur~\cite{dinur2020tight} for proving tight classical time-space tradeoffs for \emph{multiple} collision search. 

\begin{definition}[$R$-way branching program]
    An $R$-way branching program $\Bs$ with length $T$ and width $2^S$ is a layered directed graph with layers ordered $0, 1, \ldots, T$ and $2^S$ vertices in layers $1, 2, \ldots, T-1$. Layer 0 has a single vertex, called the start vertex. Layer $T$ contains (arbitrarily many) leaf vertices, each corresponding to a specific output of the branching program. All other vertices in layers $0\leq t < T$ have out-degree $R$. These edges are indexed $1,2,\ldots, R$ and each point to a vertex in layer $t+1$ (the computation is generally irreversible, so different edges may point to the same vertex).
    
    Let $f: \Us\rightarrow [R]$ be an input to the branching program. Each non-leaf vertex is associated with a query $x\in \Us$, and the outgoing edges correspond to each possible output $f(x)\in [R]$. A fixed input $f$ therefore defines a unique computation path which traverses exactly one vertex in each layer. The goal of the branching program is to compute some function depending on the input $f$.
\end{definition}

In our setting the input function $f= H$ is an oracle $H\sim \Hm$ drawn uniformly at random and $R= N$ denotes the number of possible query outcomes. Define the memory map $\mu_t$ which maps oracles $H$ to the branching program state $\mu_t(H)$ in layer $t$ that is reached after $t$ queries (in bijection with the vertices in layer $t$). The label-permutation action $H\mapsto\sigma\circ H$ on oracles induces a symmetric-group action on reachable memory states, which we denote using the same symbol:
\begin{equation}\label{eq:classical-memory-action}
    \sigma\circ\mu_t(H)\triangleq \mu_t(\sigma\circ H).
\end{equation}
The equivalence in \eqref{eq:ls-classical-symmetry} makes this well-defined: oracles reaching the same memory state still reach the same memory state after a common relabeling. This action preserves each layer.

\begin{remark}
Label symmetry is also equivalent to the existence, for every $t$, of an action of $\mathfrak{S}_N$ on reachable memory states for which the map $\mu_t$ from oracles to memory states is equivariant.
\end{remark}

\subsection{The classical matched label set}
We say a set $U\subseteq[N]$ is \emph{matched} under memory state $m_t$ if every permutation fixing the labels of $U$ pointwise fixes the memory state $m_t$. The labels seen along any execution reaching $m_t$ form a matched set of size at most $t$, since fixing these labels preserves the transcript. For $t<N/2$, the intersection of two matched sets $U,V$ of size at most $t$ is also a matched set: since $U\cup V\neq[N]$, the permutations fixing $U\cap V$ pointwise are generated by those fixing $U$ or $V$ pointwise. Consequently, there is a unique matched set of minimum cardinality (by taking successive intersections), which we denote $L(m_t) \subseteq [N]$. It is contained in every matched set of size at most $t$, so its labels have been seen along every branching program execution path reaching $m_t$, and $|L(m_t)|\leq t$.

Relabeling preserves matched sets and their sizes, so uniqueness of $L(m_t)$ gives equivariance: $L(\sigma\circ m_t)=\sigma(L(m_t))$. In particular, every permutation fixing $m_t$ preserves $L(m_t)$ as a set, while every permutation fixing $L(m_t)$ pointwise fixes $m_t$.

\begin{lemma}[Matched label set bound]\label{lem:classical-support-space}
For a fixed $t<\min\{N/2, T\}$, every reachable state $m_t$ satisfies
\begin{equation}\label{eq:classical-support-space}
    \binom{N}{|L(m_t)|}\leq2^S.
\end{equation}
Thus $|L(m_t)|\log(N/|L(m_t)|)\leq S$ whenever $|L(m_t)|>0$. In particular, $|L(m_t)|\leq 2S/\log N$ whenever $t\leq\sqrt N$.
\end{lemma}
\begin{proof}
The simplest proof uses the fact that, by equivariance, the $\mathfrak{S}_N$-orbit of $m_t$ contains unique states for all subsets of $[N]$ of size $|L(m_t)|$. However, we provide a proof which more closely aligns with the quantum matched label set proof to highlight the connections between the two.

Fix a reachable memory state $m_t$ and condition a uniformly random oracle $\Hm$ on $|L(\mu_t(\Hm))|=|L(m_t)|$. For each reachable memory state $m$ in layer $t$ satisfying $|L(m)|=|L(m_t)|$, define $p_m>0$ to be its conditional probability, and let $\ket{\phi_m}$ be the normalized uniform superposition of oracles $H$ satisfying $\mu_t(H)=m$. These vectors have disjoint supports and are therefore orthonormal.

Consider the coherent encoding of this conditioned oracle, summing over these layer-$t$ memory states:
\[
    \sum_{m}\sqrt{p_{m}}\ket{m}\ket{\phi_{m}}.
\]
Compress the oracle register and let $G$ be the resulting Gram matrix. Since the vectors $\ket{\phi_{m}}$ are orthonormal and compression is real, the canonical representation, after decompressing both registers, is
\[
    \sum_{m}\sqrt{p_{m}}\ket{\phi_{m}}\ket{\phi_{m}}.
\]
Thus measuring the two registers gives two oracles reaching the same memory state $m$, where $m$ has distribution $p_{m}$. Applying $L\circ \mu_t$ to each outcome gives the same matched set $L(m)$. The conditioning is invariant under relabeling, so equivariance of $L$ makes this common set uniform over the $\binom{N}{|L(m_t)|}$ subsets of size $|L(m_t)|$. Measuring the original oracle register prepares a pure-state ensemble on the dual register whose span has dimension $\rank(G)$. The Holevo bound (\cref{lem:holevo-bound}) and data processing therefore give
\[
    \log\binom {N}{|L(m_t)|}\leq\log\rank(G)\leq S.
\]

For $|L(m_t)|>0$, the remaining claims follow from $\binom{N}{|L(m_t)|}\geq(N/|L(m_t)|)^{|L(m_t)|}$ and $|L(m_t)|\leq t\leq\sqrt N$ in the last case; the $|L(m_t)|=0$ case is immediate.
\end{proof}

Now we consider how the set $L(\mu_t(H))$ changes over the execution of the branching program. Given oracle $H$, let $m_t$ denote the memory state $\mu_t(H)$ after $t$ queries and let $y_t$ be the next oracle answer. Every label-permutation fixing $L(m_t)\cup\{y_t\}$ pointwise also fixes the current memory state and query response. The original and relabeled executions therefore will make the same query and reach the same next state. Thus $L(m_t)\cup\{y_t\}$ is a matched set for $m_{t+1}$ of size at most $t+1$. Additionally, for $t+1<N/2$, it also contains the minimum matched set $L(m_{t+1})$. Hence
\begin{equation}\label{eq:classical-support-update}
    L(m_{t+1})\subseteq L(m_t)\cup\{y_t\}.
\end{equation}

\subsection{Comparison with an injective oracle}
We now use the memory action and matched sets $L(m_t)$ to bound the progress of the adversary per query.

We utilize the ``game-playing'' framework of \cite{BR06} to compare the branching program’s memory states with relabelings of the states reached by a collision-free oracle. Specifically, given an injective oracle $H_*:x\mapsto x$, write $m_t^*$ for the branching program memory state at time $t$ and $x_t^*$ for its query at time $t$. Denote $L_t^*\triangleq L(m_t^*)$. The response to query $x_t^*$ is $x_t^*$, so $L_t^* \subseteq \{x_0^*,\ldots,x_{t-1}^*\}$.

We say that a label-permutation $\sigma\in\mathfrak{S}_N$ \emph{extends} oracle $H$ on $L_t^*$ if
\begin{equation}
    \sigma(H_*(x)) = H(x),\quad \text{for all}\ x\in L_t^*.
\end{equation}
As long as $H$ is injective on $L_t^*$ (contains no collisions on its restriction), then there exists a $\sigma$ which extends $H$ on $L_t^*$ (that may be different at each query). We compare $H$ and $H_*$ by maintaining the correspondence $\mu_t(H)=\sigma\circ m_t^*$ throughout the execution, extending the relabeling to the next query $x_t^*$ on $H_*$ at each step. Using the Difference Lemma~\cite[Lemma~1]{Shoup2004games}, the probability the adversary can find a collision is at most the probability that the required relabeling cannot be extended at some query, plus an additive $1/N$ term accounting for a collision guessed at the output. The proof below makes this comparison explicit.

\begin{remark}
    This argument can be extended to random oracles
    $\Hm:[M]\rightarrow[N]$ where $M\geq N$ by setting $H_*$ to be
    any oracle constructed with lazy sampling which is injective on the set of queries that the branching program makes on it. Labels seen in the
    computation on $H_*$ are identified with their unique preimages
    among those queries (for $M=N$, this preimage is identical to the label itself).
\end{remark}

\begin{lemma}[Classical progress]\label{lem:classical-label-progress}
For any label-symmetric branching program with length $T<N/2$ and width $2^S$, the success probability $p$ of outputting a valid collision is bounded as
\begin{equation}\label{eq:classical-label-progress}
    p\leq\frac{1+\sum_{t<T}|L_t^*|}{N}.
\end{equation}
\end{lemma}
\begin{proof}
For a uniformly random oracle $\Hm$, let $\Es$ be the event that
\begin{equation}
    \Hm(x_t^*)\neq\Hm(c)\quad\text{for every}\ t<T\
    \text{and}\ c\in L_t^*\setminus\{x_t^*\}.
\end{equation}
The ``bad event'' of the game is therefore $\Es^c$, and a union bound gives
\begin{equation}\label{eq:classical-reference-exception}
    \Pr[\Es^c]\leq\frac{1}{N}\sum_{t<T}|L_t^*|.
\end{equation}

Fix an oracle $\Hm=H$ for which $\Es$ holds. We claim that,
for every $t\leq T$, $H$ is injective (collision-free) on
$L_t^*$ and, for every permutation $\sigma$ extending $H$ on
$L_t^*$,
\begin{equation}\label{eq:classical-reference-coupling}
    \mu_t(H)=\sigma\circ m_t^*.
\end{equation}
We prove this claim via induction. At time $t=0$ the state is fixed and the matched set is empty. Suppose the claim holds at time $t$. The event $\Es$ and the induction hypothesis make $H$ injective on $L_t^*\cup\{x_t^*\}$. Construct a permutation
$\sigma_t$ extending $H$ on $L_t^*\cup\{x_t^*\}$, which exists
since $H$ is injective on $L_t^*\cup\{x_t^*\}$. Then $\mu_t(\sigma_t^{-1}\circ H)=m_t^*$ by the induction hypothesis and \eqref{eq:classical-memory-action}. This execution makes the query $x_t^*$ and receives the answer $x_t^*$, since $\sigma_t^{-1}(H(x_t^*))=x_t^*$. Hence
\begin{equation}
    \mu_{t+1}(H)=\sigma_t\circ\big(\mu_{t+1}(\sigma_t^{-1}\circ H)\big)=\sigma_t\circ m_{t+1}^*.
\end{equation}
By \eqref{eq:classical-support-update}, $L_{t+1}^*\subseteq L_t^*\cup\{x_t^*\}$. Thus $H$ is injective on the next matched set, and every extension of its restriction acts identically on $m_{t+1}^*$, completing the induction.

For each injective map $h:L_T^*\to[N]$, choose a permutation $\sigma_T$ extending $h$. The output prescribed by $\sigma_T\circ m_T^*$ depends only on $h$, since all extensions act identically on $m_T^*$. By the claim, this is the actual output whenever $\Es$ holds and $\Hm|_{L_T^*}=h$.

Condition only on $\Hm|_{L_T^*}=h$. The prescribed output is fixed, and the oracle values outside $L_T^*$ remain independent and uniform. If both inputs lie in $L_T^*$, they cannot collide; otherwise, their collision probability is $1/N$. The prescribed output also has to report their common label, so it satisfies the success condition with probability at most $1/N$. On $\Es$ the restriction is injective and the prescribed output is the actual output, so averaging gives
\[
    \Pr[\mathrm{success}\cap\mathcal E]\leq\frac{1}{N}.
\]
Together with \eqref{eq:classical-reference-exception}, this proves the lemma.
\end{proof}

\subsection{Completing the classical time-space tradeoff}

Now we prove the full $TS=\Omega(N\log N)$ time-space tradeoff for deterministic label-symmetric classical algorithms.

\begin{proof}[Proof of \cref{thm:classical-label-tradeoff}]
If $T\geq N/2$, the result follows from $S\geq\log N$. Suppose $T<N/2$. By \cref{lem:classical-label-progress}, $\sum_{t<T}|L_t^*|\geq 2N/3-1>0$. This sum has $T$ terms, and $|L_t^*|\leq t<T<N/2$ for every $t$, so \cref{lem:classical-support-space} applies and gives
\[
    T\max_{t<T}|L_t^*|\geq \frac{2N}{3}-1\geq \frac{N}{2},
    \qquad S\geq \big(\max_{t<T}|L_t^*|\big)\log\left(\frac{N}{\max_{t<T}|L_t^*|}\right).
\]
If $\max_{t<T}|L_t^*|\leq\sqrt N$, then $\log(N/(\max_{t<T}|L_t^*|))\geq\log N/2$, so
$TS\geq N\log N/4$. If $\max_{t<T}|L_t^*|>\sqrt N$, then $T\geq\max_{t<T}|L_t^*|>\sqrt N$. Since $\binom{N}{k}$ increases for $k\leq N/2$, the space bound gives
\[
    S\geq\log\binom N{\lfloor\sqrt N\rfloor}=\Omega(\sqrt N\log N).
\]
Thus $TS=\Omega(N\log N)$ in this case as well.
\end{proof}

%% file: citation.bib
@article{vanoorschot1999,
  title = {Parallel {{Collision Search}} with {{Cryptanalytic Applications}}},
  author = {Van Oorschot, Paul C. and Wiener, Michael J.},
  year = {1999},
  month = {1},
  journal = {Journal of Cryptology},
  volume = {12},
  number = {1},
  pages = {1--28},
  issn = {0933-2790, 1432-1378},
  doi = {10.1007/PL00003816}
}

@inproceedings{chung2021compressed,
  title={On the compressed-oracle technique, and post-quantum security of proofs of sequential work},
  author={Chung, Kai-Min and Fehr, Serge and Huang, Yu-Hsuan and Liao, Tai-Ning},
  booktitle={Annual International Conference on the Theory and Applications of Cryptographic Techniques},
  pages={598--629},
  year={2021},
  organization={Springer}
}

@inbook{Brassard_1998, 
title={Quantum cryptanalysis of hash and claw-free functions: Invited paper},
ISBN={9783540697152},
ISSN={1611-3349},
DOI={10.1007/bfb0054319},
booktitle={LATIN'98: Theoretical Informatics},
publisher={Springer Berlin Heidelberg},
author={Brassard, Gilles and H{\o}yer, Peter and Tapp, Alain},
year={1998},
pages={163–169} 
}

@article{Zhandry_2015,
author = {Zhandry, Mark},
title = {A note on the quantum collision and set equality problems},
year = {2015},
issue_date = {May 2015},
publisher = {Rinton Press, Incorporated},
address = {Paramus, NJ},
volume = {15},
number = {7–8},
issn = {1533-7146},
journal = {Quantum Info. Comput.},
month = may,
pages = {557–567},
numpages = {11}
}

@inproceedings{dinur2020tight,
  title={Tight time-space lower bounds for finding multiple collision pairs and their applications},
  author={Dinur, Itai},
  booktitle={Annual International Conference on the Theory and Applications of Cryptographic Techniques},
  pages={405--434},
  year={2020},
  organization={Springer}
}

@article{hamoudi2023quantum,
  title={Quantum time--space tradeoff for finding multiple collision pairs},
  author={Hamoudi, Yassine and Magniez, Fr{\'e}d{\'e}ric},
  journal={ACM Transactions on Computation Theory},
  volume={15},
  number={1-2},
  pages={1--22},
  year={2023},
  publisher={ACM New York, NY, USA}
}

@inproceedings{zhandry2019record,
  title={How to record quantum queries, and applications to quantum indifferentiability},
  author={Zhandry, Mark},
  booktitle={Annual International Cryptology Conference},
  pages={239--268},
  year={2019},
  organization={Springer}
}

@misc{hamoudi2025introduction,
      title={A Brief Introduction to Quantum Query Complexity}, 
      author={Yassine Hamoudi},
      year={2025},
      eprint={2508.08852},
      archivePrefix={arXiv},
      primaryClass={quant-ph}
}

@inproceedings{ambainis2000quantum,
  title={Quantum lower bounds by quantum arguments},
  author={Ambainis, Andris},
  booktitle={Proceedings of the 32nd Annual ACM Symposium on Theory of Computing},
  pages={636--643},
  year={2000},
  series= {STOC '00},
}

@article{Bennett_1997,
   title={Strengths and Weaknesses of Quantum Computing},
   volume={26},
   ISSN={1095-7111},
   DOI={10.1137/s0097539796300933},
   number={5},
   journal={SIAM Journal on Computing},
   publisher={Society for Industrial & Applied Mathematics (SIAM)},
   author={Bennett, Charles H. and Bernstein, Ethan and Brassard, Gilles and Vazirani, Umesh},
   year={1997},
   month=oct, pages={1510–1523} }

@inproceedings{BR93,
author = {Bellare, Mihir and Rogaway, Phillip},
title = {Random oracles are practical: a paradigm for designing efficient protocols},
year = {1993},
isbn = {0897916298},
publisher = {Association for Computing Machinery},
address = {New York, NY, USA},
doi = {10.1145/168588.168596},
booktitle = {Proceedings of the 1st ACM Conference on Computer and Communications Security},
pages = {62–73},
numpages = {12},
location = {Fairfax, Virginia, USA},
series = {CCS '93}
}

@inbook{Boneh_2011,
   title={Random Oracles in a Quantum World},
   ISBN={9783642253850},
   ISSN={1611-3349},
   DOI={10.1007/978-3-642-25385-0_3},
   booktitle={Advances in Cryptology – ASIACRYPT 2011},
   publisher={Springer Berlin Heidelberg},
   author={Boneh, Dan and Dagdelen, Özgür and Fischlin, Marc and Lehmann, Anja and Schaffner, Christian and Zhandry, Mark},
   year={2011},
   pages={41–69} }

@article{klauck2007quantum,
  title={Quantum and classical strong direct product theorems and optimal time-space tradeoffs},
  author={Klauck, Hartmut and {\v{S}}palek, Robert and De Wolf, Ronald},
  journal={SIAM Journal on Computing},
  volume={36},
  number={5},
  pages={1472--1493},
  year={2007},
  publisher={SIAM}
}

@INPROCEEDINGS{BSS_2003,
  author={Barnum, H. and Saks, M. and Szegedy, M.},
  booktitle={18th IEEE Annual Conference on Computational Complexity, 2003. Proceedings.}, 
  title={Quantum query complexity and semi-definite programming}, 
  year={2003},
  volume={},
  number={},
  pages={179-193},
  doi={10.1109/CCC.2003.1214419}}

@article{bennett1989time,
  title={Time/space trade-offs for reversible computation},
  author={Bennett, Charles H},
  journal={SIAM Journal on Computing},
  volume={18},
  number={4},
  pages={766--776},
  year={1989},
  publisher={SIAM}
}

@article{aaronsonshi2004,
author = {Aaronson, Scott and Shi, Yaoyun},
title = {Quantum lower bounds for the collision and the element distinctness problems},
year = {2004},
issue_date = {July 2004},
publisher = {Association for Computing Machinery},
address = {New York, NY, USA},
volume = {51},
number = {4},
issn = {0004-5411},
doi = {10.1145/1008731.1008735},
journal = {Journal of the ACM},
month = {7},
pages = {595–605},
numpages = {11}
}

@article{pollardrho1975,
  title = {A Monte Carlo Method for Factorization},
  author = {Pollard, J. M.},
  year = {1975},
  month = {9},
  journal = {BIT Numerical Mathematics},
  volume = {15},
  number = {3},
  pages = {331--334},
  issn = {1572-9125},
  doi = {10.1007/BF01933667}
}

@misc{magniez2026label,
      title={Tight Time-Space Lower Bounds for Collision Finding and Element Distinctness under Label Symmetry}, 
      author={Frédéric Magniez and Sebastian Zur},
      year={2026},
      eprint={2609.10808v2},
      archivePrefix={arXiv},
      primaryClass={quant-ph},
}

@article{HW94,
author = {Paul Hausladen and William K. Wootters},
title = {A ‘Pretty Good’ Measurement for Distinguishing Quantum States},
journal = {Journal of Modern Optics},
volume = {41},
number = {12},
pages = {2385--2390},
year = {1994},
publisher = {Taylor \& Francis},
doi = {10.1080/09500349414552221}
}

@ARTICLE{EldarForney2001,
  author={Eldar, Y.C. and Forney, G.D.},
  journal={IEEE Transactions on Information Theory}, 
  title={On quantum detection and the square-root measurement}, 
  year={2001},
  volume={47},
  number={3},
  pages={858-872},
  doi={10.1109/18.915636}}

@book{Watrous_2018, place={Cambridge}, title={The Theory of Quantum Information}, publisher={Cambridge University Press}, author={Watrous, John}, year={2018}}

@inproceedings{Hao2026short,
author = {Hao, Zihan and Huang, Zikuan and Liu, Qipeng},
title = {On the Need for (Quantum) Memory with Short Outputs},
year = {2026},
isbn = {9798400725364},
publisher = {Association for Computing Machinery},
address = {New York, NY, USA},
doi = {10.1145/3798129.3800896},
booktitle = {Proceedings of the 58th Annual ACM Symposium on Theory of Computing},
pages = {1901–1912},
numpages = {12},
location = {Salt Lake City, UT, USA},
series = {STOC '26}
}

@article{Aaronson2021open,
author = {Aaronson, Scott},
title = {Open Problems Related to Quantum Query Complexity},
year = {2021},
issue_date = {December 2021},
publisher = {Association for Computing Machinery},
address = {New York, NY, USA},
volume = {2},
number = {4},
doi = {10.1145/3488559},
journal = {ACM Transactions on Quantum Computing},
month = {12},
articleno = {14},
numpages = {9}
}

@article{beam2026matrix,
author = {Beame, Paul and Kornerup, Niels and Whitmeyer, Michael},
title = {Quantum Time-Space Tradeoffs for Matrix Problems},
journal = {SIAM Journal on Computing},
volume = {55},
number = {3},
pages = {469-519},
year = {2026},
doi = {10.1137/24M1710164}
}

@INPROCEEDINGS{chung2020function,
  author={Chung, Kai-Min and Guo, Siyao and Liu, Qipeng and Qian, Luowen},
  booktitle={2020 IEEE 61st Annual Symposium on Foundations of Computer Science (FOCS)}, 
  title={Tight Quantum Time-Space Tradeoffs for Function Inversion}, 
  year={2020},
  volume={},
  number={},
  pages={673-684},
  doi={10.1109/FOCS46700.2020.00068}}

@inproceedings{liu2023nonuniformity,
author = {Liu, Qipeng},
title = {Non-uniformity and Quantum Advice in the Quantum Random Oracle Model},
year = {2023},
isbn = {978-3-031-30544-3},
publisher = {Springer-Verlag},
address = {Berlin, Heidelberg},
doi = {10.1007/978-3-031-30545-0_5},
booktitle = {Advances in Cryptology – EUROCRYPT 2023: 42nd Annual International Conference on the Theory and Applications of Cryptographic Techniques, Lyon, France, April 23-27, 2023, Proceedings, Part I},
pages = {117–143},
numpages = {27}
}

@InProceedings{Banegas2017search,
author="Banegas, Gustavo and Bernstein, Daniel J.",
editor="Adams, Carlisle and Camenisch, Jan",
title="Low-Communication Parallel Quantum Multi-Target Preimage Search",
booktitle="Selected Areas in Cryptography -- SAC 2017",
year="2018",
publisher="Springer International Publishing",
pages="325--335"
}

@inproceedings{jaques2020lowgate,
author = {Jaques, Samuel and Schrottenloher, Andr{\'e}},
title = {Low-Gate Quantum Golden Collision Finding},
year = {2020},
isbn = {978-3-030-81651-3},
publisher = {Springer-Verlag},
address = {Berlin, Heidelberg},
doi = {10.1007/978-3-030-81652-0_13},
booktitle = {Selected Areas in Cryptography: 27th International Conference, Halifax, NS, Canada (Virtual Event), October 21-23, 2020, Revised Selected Papers},
pages = {329–359},
numpages = {31}
}

@article{beals01,
  author    = {Beals, Robert and Buhrman, Harry and Cleve, Richard and Mosca, Michele and de Wolf, Ronald},
  title     = {Quantum Lower Bounds by Polynomials},
  journal   = {Journal of the ACM},
  volume    = {48},
  number    = {4},
  year      = {2001},
  pages     = {778--797},
  doi       = {10.1145/502090.502097}
}

@article{montanaro2007pgm,
  title = {On the {{Distinguishability}} of {{Random Quantum States}}},
  author = {Montanaro, Ashley},
  year = {2007},
  month = {8},
  journal = {Communications in Mathematical Physics},
  volume = {273},
  number = {3},
  pages = {619--636},
  issn = {1432-0916}
}

@inproceedings{HLS07,
  author    = {Peter H{\o}yer and Troy Lee and Robert {\v{S}}palek},
  title     = {Negative weights make adversaries stronger},
  booktitle = {Proceedings of the 39th Annual ACM Symposium on Theory of Computing},
  series    = {STOC '07},
  pages     = {526--535},
  address   = {San Diego, California, USA},
  year      = {2007},
  doi       = {10.1145/1250790.1250867}
}

@article{reichardt2011span,
author = {Reichardt, Ben W.},
title = {Span Programs are Equivalent to Quantum Query Algorithms},
journal = {SIAM Journal on Computing},
volume = {43},
number = {3},
pages = {1206-1219},
year = {2014},
doi = {10.1137/100792640}
}

@article{barnumknill2002,
  title = {Reversing Quantum Dynamics with Near-Optimal Quantum and Classical Fidelity},
  author = {Barnum, H. and Knill, E.},
  year = {2002},
  month = {5},
  journal = {Journal of Mathematical Physics},
  volume = {43},
  number = {5},
  pages = {2097--2106},
  issn = {0022-2488},
  doi = {10.1063/1.1459754}
}

@InProceedings{BR06,
    author="Bellare, Mihir
    and Rogaway, Phillip",
    editor="Vaudenay, Serge",
    title="The Security of Triple Encryption and a Framework for Code-Based Game-Playing Proofs",
    booktitle="Advances in Cryptology - EUROCRYPT 2006",
    year="2006",
    publisher="Springer Berlin Heidelberg",
    address="Berlin, Heidelberg",
    pages="409--426",
    isbn="978-3-540-34547-3"
}

@article{BorodinCook1982,
    author = {Borodin, A. and Cook, S.},
    title = {A Time-Space Tradeoff for Sorting on a General Sequential Model of Computation},
    journal = {SIAM Journal on Computing},
    volume = {11},
    number = {2},
    pages = {287-297},
    year = {1982},
    doi = {10.1137/0211022}
}

@misc{Shoup2004games,
      author = {Victor Shoup},
      title = {Sequences of games: a tool for taming complexity in security proofs},
      howpublished = {Cryptology {ePrint} Archive, Paper 2004/332},
      year = {2004},
      url = {https://eprint.iacr.org/2004/332}
}

@misc{gao2026query,
      title={Quantum Query Advantage Requires Space}, 
      author={Minbo Gao and Zhengfeng Ji and Ziyi Xie},
      year={2026},
      eprint={2609.29497},
      archivePrefix={arXiv},
      primaryClass={quant-ph}
}

@inproceedings{Lee2011Quantum,
  author    = {Troy Lee and
               Rajat Mittal and
               Ben W. Reichardt and
               Robert \v{S}palek and
               Mario Szegedy},
  title     = {Quantum query complexity of state conversion},
  booktitle = {Proceedings of the 52nd Annual IEEE Symposium on Foundations of Computer Science (FOCS)},
  pages     = {344--353},
  year      = {2011},
  doi       = {10.1109/FOCS.2011.75}
}

@inproceedings{Bernstein2009Cost,
  author    = {Daniel J. Bernstein},
  title     = {Cost analysis of hash collisions: Will quantum computers make {SHARCS} obsolete?},
  booktitle = {Proceedings of the 4th Workshop on Special-purpose Hardware for Attacking Cryptographic Systems ({SHARCS} '09)},
  pages     = {105--116},
  year      = {2009},
  address   = {Lausanne, Switzerland},
  month     = {9}
}

@InProceedings{CNS2017,
author="Chailloux, Andr{\'e}
and Naya-Plasencia, Mar{\'i}a
and Schrottenloher, Andr{\'e}",
editor="Takagi, Tsuyoshi
and Peyrin, Thomas",
title="An Efficient Quantum Collision Search Algorithm and Implications on Symmetric Cryptography",
booktitle="Advances in Cryptology -- ASIACRYPT 2017",
year="2017",
publisher="Springer International Publishing",
address="Cham",
pages="211--240"
}

@article{Grover2004collision,
author = {Grover, Lov and Rudolph, Terry},
title = {How significant are the known collision and element distinctness quantum algorithms},
year = {2004},
publisher = {Rinton Press, Incorporated},
address = {Paramus, NJ},
volume = {4},
number = {3},
issn = {1533-7146},
journal = {Quantum Info. Comput.},
month = {5},
pages = {201–206},
numpages = {6}
}

@InProceedings{GR22,
  author =	{Girish, Uma and Raz, Ran},
  title =	{{Eliminating Intermediate Measurements Using Pseudorandom Generators}},
  booktitle =	{13th Innovations in Theoretical Computer Science Conference (ITCS 2022)},
  pages =	{76:1--76:18},
  series =	{Leibniz International Proceedings in Informatics (LIPIcs)},
  ISBN =	{978-3-95977-217-4},
  ISSN =	{1868-8969},
  year =	{2022},
  volume =	{215},
  editor =	{Braverman, Mark},
  publisher =	{Schloss Dagstuhl -- Leibniz-Zentrum f{\"u}r Informatik},
  address =	{Dagstuhl, Germany},
  URN =		{urn:nbn:de:0030-drops-156726},
  doi =		{10.4230/LIPIcs.ITCS.2022.76}
}

@InProceedings{Zha24,
  author =	{Zhandry, Mark},
  title =	{{The Space-Time Cost of Purifying Quantum Computations}},
  booktitle =	{15th Innovations in Theoretical Computer Science Conference (ITCS 2024)},
  pages =	{102:1--102:22},
  series =	{Leibniz International Proceedings in Informatics (LIPIcs)},
  ISBN =	{978-3-95977-309-6},
  ISSN =	{1868-8969},
  year =	{2024},
  volume =	{287},
  editor =	{Guruswami, Venkatesan},
  publisher =	{Schloss Dagstuhl -- Leibniz-Zentrum f{\"u}r Informatik},
  address =	{Dagstuhl, Germany},
  URN =		{urn:nbn:de:0030-drops-196304},
  doi =		{10.4230/LIPIcs.ITCS.2024.102}
}
